\documentclass[twocolumn]{aastex7}
\usepackage{nicefrac}
\usepackage{amsmath}

\usepackage{todonotes}

\begin{document}

\title{Massive Interacting Binaries Enhance Feedback in Star-Forming Regions}

\correspondingauthor{Claude Cournoyer-Cloutier}
\email{cournoyc@mcmaster.ca}

\author[0000-0002-6116-1014]{Claude Cournoyer-Cloutier}
\affiliation{Department of Physics and Astronomy, McMaster University, 1280 Main Street West, Hamilton, ON, L8S 4M1, Canada}
\email{cournoyc@mcmaster.ca}

\author[0000-0003-3479-4606]{Eric P. Andersson}
\affiliation{Department of Astrophysics, American Museum of Natural History, 200 Central Park West, New York, NY 10024-5102, USA}
\email{eandersson@amnh.org}

\author[0000-0002-6593-3800]{Sabrina M. Appel}
\altaffiliation{NSF Astronomy \& Astrophysics Postdoctoral Fellow}
\affiliation{Department of Astrophysics, American Museum of Natural History, 200 Central Park West, New York, NY 10024-5102, USA}
\email{sappel@amnh.org}

\author[0000-0003-2166-1935]{Natalia Lahén}
\affiliation{Max Planck Institute for Astrophysics, Karl-Schwarzschild-Straße 1, 85748 Garching bei München, Germany}
\email{nlahen@MPA-Garching.MPG.DE}

\author[0000-0001-5972-137X]{Brooke Polak}
\affiliation{Department of Astrophysics, American Museum of Natural History, 200 Central Park West, New York, NY 10024-5102, USA}
\email{bpolak@amnh.org}

\author[0000-0001-8789-2571]{Antti Rantala}
\affiliation{Max Planck Institute for Astrophysics, Karl-Schwarzschild-Straße 1, 85748 Garching bei München, Germany}
\email{anttiran@mpa-garching.mpg.de}

\author[0000-0002-2998-7940]{Silvia Toonen}
\affiliation{Anton Pannekoek Institute for Astronomy, University of Amsterdam, Science Park 904, 1098 XH Amsterdam, The Netherlands}
\email{S.G.M.Toonen@uva.nl}

\author[0000-0003-3551-5090]{Alison Sills}
\affiliation{Department of Physics and Astronomy, McMaster University, 1280 Main Street West, Hamilton, ON, L8S 4M1, Canada}
\email{asills@mcmaster.ca}

\author[0000-0003-3688-5798]{Steven Rieder}
\affiliation{Anton Pannekoek Institute for Astronomy, University of Amsterdam, Science Park 904, 1098 XH Amsterdam, The Netherlands}
\email{s.rieder@uva.nl}

\author[0000-0001-5839-0302]{Simon Portegies Zwart}
\affiliation{Sterrewacht Leiden, Leiden University, Einsteinweg 55, 2333CC Leiden, The Netherlands}
\email{spz@strw.leidenuniv.nl}

\author[0000-0003-0064-4060]{Mordecai-Mark Mac Low}
\affiliation{Department of Astrophysics, American Museum of Natural History, 200 Central Park West, New York, NY 10024-5102, USA}
\email{mordecai@amnh.org}

\author[0000-0001-8762-5772]{William E. Harris}
\affiliation{Department of Physics and Astronomy, McMaster University, 1280 Main Street West, Hamilton, ON, L8S 4M1, Canada}
\email{harrisw@mcmaster.ca}



\begin{abstract}
We present a new framework to incorporate feedback from massive interacting binaries in simulations of star cluster formation. Our new feedback model adds binary stellar evolution to the cluster formation code \textsc{Torch}, and couples it in \textsc{Amuse} to the pre-existing modules for collisional stellar dynamics, magnetohydrodynamics, and mechanical and radiative feedback. Our model accounts for the effects of mass transfer on the stars' mass loss rates, their radiation spectra, and the timing of core-collapse supernovae. It also injects mass lost through non-conservative mass transfer and common envelope ejection into the interstellar medium. We demonstrate the use of our feedback model through simulations of isolated binaries in a gaseous medium, and of embedded clusters of massive binaries. Feedback from interacting binaries efficiently couples with the surrounding interstellar medium. It increases the size of \ion{H}{2} regions, increases the kinetic and thermal energy of the gas, and increases the pressure within \ion{H}{2} regions compared to models that use single star stellar evolution. Those differences arise from the ionizing radiation, which increases by three orders of magnitude, resulting in \ion{H}{2} regions that expand due to thermal pressure rather than radiation pressure. The effects of stellar dynamics and the gravitational potential of the background gas cause the evolution of individual binaries to deviate from the predictions made by secular evolution, impacting the subsequent feedback from the binary. We conclude that massive interacting binaries are an important source of feedback in cluster-forming regions, and must be considered when studying the emerging timescales of young star clusters.
\end{abstract}



\setcounter{footnote}{0} 
\section{Introduction} \label{sec:intro}
Massive stars play an important role in the evolution of star-forming 
galaxies by injecting energy and momentum into the interstellar medium (ISM) via stellar feedback. 
O-type stars are the dominant source of feedback in regions forming massive star clusters, rather than protostellar feedback from lower-mass stars~\citep[][]{Matzner2015, Plunkett2015}. Different feedback mechanisms dominate at different times: core-collapse supernova (SN) feedback, which injects mass and energy into the ISM after $\gtrsim 3$ Myr, and pre-SN feedback, which acts from the time of star formation. Massive stars regulate star formation by providing mechanical and radiative feedback to the surrounding ISM through stellar winds~\citep[][]{Rogers2013, Wareing2017, Geen2021, Lancaster2021, Lancaster2024}, radiation~\citep[][]{Howard2017, Barnes2020, Menon2023}, and core-collapse SNe~\citep[][]{Walch2015, Koertgen2016, Lucas2020}. Recent observations of young massive star clusters (YMCs) suggest that pre-SN feedback is sufficient to interrupt star formation and remove the gas from the cluster-forming region~\citep[]{Chevance2022, Hannon2022, Deshmukh2024}. However, the masses and radii of YMCs observed in local starburst galaxies~\citep[][]{Leroy2018, He2022, Levy2024} and at high $z$~\citep[][]{Vanzella2023, Adamo2024} suggest that that they are dense enough to prevent radiative and wind feedback from halting star formation~\citep[][]{Krumholz2019}. More work on the effects of pre-SN stellar feedback on the ISM at the scale of individual YMCs is necessary to resolve this tension. 

Milky Way O stars have a multiplicity fraction $\gtrsim 95\%$~\citep{Moe2017, Offner2023}. Observations reveal high binary fractions for massive stars in local YMCs~\citep[][]{Almeida2017, Ritchie2022, Clark2023}; those binaries can survive in denser, more massive clusters that mimic those forming in starburst galaxies~\citep{Cournoyer-Cloutier2024b}.
%
%
At least 70\% of all O stars have a companion within 10 au; as a consequence, 70\%~\citep{Sana2012} to $\gtrsim$~90\%~\citep{Moe2017} of O stars undergo a pre-SN mass transfer (MT) episode. 
This has profound implications for the timing and strength of the feedback injected into the ISM.

Stars in close binaries undergo MT through Roche lobe overflow (RLOF) which is triggered once the radius of a star exceeds its Roche lobe~\citep[][]{Paczynski1971, Eggleton1983}. The donor
loses material from its envelope, which is accreted by the companion or lost from the system; those outcomes are respectively known as conservative and non-conservative MT~\citep[][]{Soberman1997}. The radii, temperatures, and luminosities of the stars change due to MT, modifying their radiative and mechanical feedback~\citep[][and references therein]{Marchant2024}. 
MT via RLOF is commonly labeled based on the donor's evolutionary stage: Case A for MT while the donor is on the main sequence (MS), and Cases B and C for to MT on the post-MS before and after the end of core helium burning~\citep{Kippenhahn1967, Lauterborn1970}. 
For massive stars at solar metallicity, Case A and Case B MT can occur for orbital periods of $\lesssim$~1000 and $\lesssim$~5000 days~\citep[][]{Moe2017}.

Stars in close binaries may also undergo a common envelope (CE) phase~\citep[][]{Paczynski1976, Ivanova2013, Roepke2023}, during which the donor's envelope engulfs the companion and is subsequently ejected from the system. Both CE ejection and non-conservative MT result in very high instantaneous mass loss rates: CE ejection unbinds the primary's envelope within $\lesssim$1000 years~\citep[based on its dynamical timescale,][]{Ivanova2015} while rapid MT leads to mass loss rates $\lesssim10^{-2}$~M$_{\odot}$ yr$^{-1}$~\citep[assuming mass transfer on the thermal timescale,][]{deMink2007},
both several orders of magnitude larger than wind mass loss rates~\citep[$\lesssim$ 10$^{-4}$ M$_{\odot}$ yr$^{-1}$,][]{Vink2022}. 

Non-conservative MT and CE ejection in massive binaries
increase the amount of pre-SN ejecta from a stellar population compared to a population of single stars~\citep[e.g.][]{Farmer2023, Nguyen2024}. 
MT also increases the amount of far ultraviolet~\citep[FUV,][]{Goetberg2018} and ionizing radiation~\citep{Goetberg2020} from massive stars, by rejuvenating accretors and exposing the hot 
cores of stripped stars. 
%
%
Conservative and non-conservative MT in massive binaries change the timing of core-collapse SNe~\citep{PortegiesZwart1996, Zapartas2017}: they can either hasten or delay SNe in individual systems, change a star's explodability~\citep[e.g.][]{Woosley2019, Antoniadis2022, Laplace2025}, and change the timing of the first SN in a YMC.
Any comprehensive model of stellar feedback should therefore account for the binary evolution of the population of O stars.

YMCs are dynamically rich environments. Few-body interactions may disrupt binaries~\citep[][]{Heggie1975, Hills1975}. 
Repeated encounters tend to decrease the semimajor axis~\citep[][]{Heggie1975, Hills1975} and increase the eccentricity~\citep[][]{Heggie1996} of a binary, changing the future MT episodes of the system beyond what is predicted by standalone binary stellar evolution. Few-body interactions also result in exchanges~\citep[][]{Sigurdsson1993}, and massive stars can pair up with a new companion that has already undergone MT. Those interactions are in turn driven by YMC formation: hierarchical cluster assembly leads to bursts of few-body interactions~\citep[][]{Fujii2012, Rantala2024}
due to violent relaxation~\citep{Lynden-Bell1967}, which are strengthened in the presence of primordial binaries and background gas driving the assembly process~\citep{Cournoyer-Cloutier2024a}. A comprehensive treatment of feedback from massive binaries must account for stellar dynamics, hydrodynamics, and binary stellar evolution simultaneously. 

%
Previous star-by-star hydrodynamical models of cluster formation have adopted stellar evolution and stellar feedback schemes based on single star models, although some have included dynamical binary formation through collisional stellar dynamics~\citep[e.g.,][]{Wall2019, Fujii2021, Polak2024b, Lahen2025} or primordial binaries~\citep{Cournoyer-Cloutier2021, Cournoyer-Cloutier2023, Cournoyer-Cloutier2024b}. 
Cluster evolution codes include both primordial binaries and binary stellar evolution, but no hydrodynamics, and therefore no stellar feedback~\citep[e.g.,][]{PortegiesZwart1999, Chatterjee2010, Hypki2013, Wang2015, Rodriguez2022, Rantala2025}. 
Investigating the effects of coupled binary stellar evolution and dynamics on the feedback budget of YMCs is needed for understanding how stellar feedback halts star formation in massive cluster-forming regions. 


In this paper, we present a new framework for feedback from massive interacting binaries in star-forming regions, and demonstrate the importance of feedback from massive interacting binaries on the nearby ISM. The model accounts for the effects of binary evolution on the radiative and mechanical feedback from massive stars, allows for binaries to be modified through the effects of stellar dynamics, and models the binaries alongside the background gas. 
This framework is implemented within the \textsc{Torch} cluster formation model~\citep{Wall2019}, which couples stellar dynamics to star and binary~\citep{Cournoyer-Cloutier2021} formation, and stellar evolution. The coupling of binary evolution to stellar dynamics and hydrodynamics is described in Section~\ref{sec:coupling}. Section~\ref{sec:MT} gives examples of conservative and non-conservative mass transfer. We discuss a suite of example simulations of compact clusters of massive binaries in Section~\ref{sec:cluster}, followed by a discussion and summary in Section~\ref{sec:discussion}. 

\section{Coupling binary evolution with stellar dynamics and hydrodynamics}\label{sec:coupling}

Binary evolution can change the binary's orbit,
trigger mass loss, and change the stars' radiation spectrum due to mass transfer. The effects of gravity from nearby stars and gas can also modify the orbits of the binaries, and in turn influence binary evolution. It is thus crucial to ensure that the coupling between star and binary evolution, stellar dynamics, stellar feedback, and hydrodynamics
allows information to be propagated correctly between the different codes.

\setcounter{footnote}{0}
\subsection{\textsc{Torch}}

\begin{figure}[tb!]
    \centering
    \includegraphics[width=\linewidth, clip=True, trim=2.5cm 9cm 2.5cm 1cm]{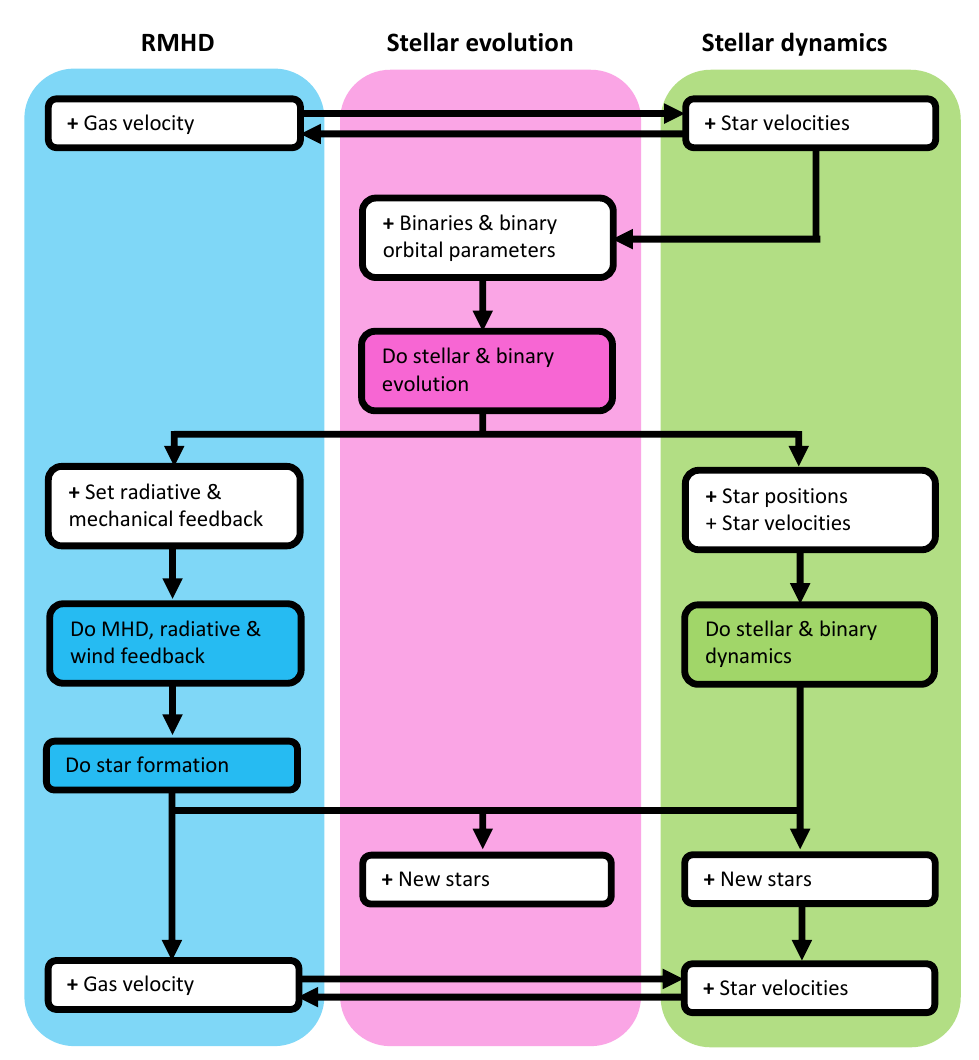}
    \caption{Flowchart showing the order of operations and information passed between the codes handling RMHD (left, \textsc{Flash}), stellar and binary evolution (center, \textsc{SeBa}), and stellar dynamics (right, \textsc{PeTar}), for one \textsc{Torch} time step. The boxes with a white background denote updates from one code to another, while the boxes with a darker background correspond to operations done within one code.}
    \label{fig:flowchart}
\end{figure}

\textsc{Torch}~\citep{Wall2019}\footnote{\url{https://bitbucket.org/torch-sf/torch/commits/tag/interacting-binaries-v1.0}, branch used in this paper}\textsuperscript{,}\footnote{\url{https://bitbucket.org/torch-sf/torch/commits/tag/torch-v2.0}, current stable version} uses the \textsc{Amuse} framework~\citep{PortegiesZwart2009, Pelupessy2013, PortegiesZwart2013, PortegiesZwart2019, 
amuse}\footnote{\url{https://github.com/amusecode/amuse}, commit \texttt{aea5b55}} to couple radiation magnetohydrodynamics (RMHD) to collisional stellar dynamics and stellar evolution. 
\textsc{Torch} is designed as a star cluster formation code, and includes star~\citep{Wall2019} and binary~\citep{Cournoyer-Cloutier2021} formation.  
Stellar evolution is handled by \textsc{SeBa}~\citep{PortegiesZwart1996, Nelemans2001, Toonen2012} while mechanical and radiative feedback from massive stars is handled directly in the RMHD code \textsc{Flash}~\citep{Fryxell2000, Dubey2014}. In this paper, we present the implementation of binary stellar evolution and feedback from interacting binary stars in \textsc{Torch}.

The order of operations within one code time step are presented in Figure~\ref{fig:flowchart}. The gravitational interactions between the gas and the stars are handled with a leapfrog scheme based on \textsc{Bridge}~\citep[][]{Fujii2007}, which takes a kick-drift-kick approach described in~\citet{Wall2019}. Each step begins and ends with a \textit{kick} updating the gas velocity from the stars and vice-versa; the corresponding acceleration is calculated with a multi-grid solver~\citep{Ricker2008}. 
Binaries are identified every time step, after the first kick, as described in Section~\ref{sec:binaryID}. Once the information about the binaries has been passed to the stellar and binary evolution code, the single stars and the binaries are evolved by a time step $\Delta t$. The stellar and binary evolution in \textsc{SeBa} is presented in Section~\ref{sec:seba}. The output from \textsc{SeBa} is used to update the stars' positions and velocities, and to set the radiative and mechanical feedback properties. 
We describe in Sections~\ref{sec:detection} and~\ref{sec:mloss} how we adapt the feedback routine to account for binary evolution. After the information from stellar and binary evolution has been passed to the RMHD and stellar dynamics codes, those codes evolve in parallel to complete the \textit{drift} part of the \textsc{Torch} step; the relevant information for the RMHD and stellar dynamics methods is presented in Section~\ref{sec:flash} and~\ref{sec:petar}. The simulation time step $\Delta t$ is set from the minimum time step of the magnetohydrodynamics (MHD) and stellar dynamics solvers; the radiation and stellar and binary evolution solvers can adopt shorter time steps and subcycle over $\Delta t$ as needed.

\subsection{Radiation magnetohydrodynamics}\label{sec:flash}
We use the adaptive mesh refinement code \textsc{Flash} to evolve MHD and radiation.
We use the HLLD Riemann solver~\citep{Miyoshi2005} with a multi-grid solver~\citep{Ricker2008} for the gas self-gravity. We refine the grid with the second-derivative criterion~\citep[][]{Lohner1987} 
for temperature and pressure, and require the Jeans length to be resolved by a minimum of 12 cells~\citep{Heitsch2001, Federrath2010}. 

Radiation is handled through the ray-tracing module \textsc{Fervent}~\citep{Baczynski2015}. For numerical stability, we allow the ionization fraction to change by a maximum of 10\% per radiation step, and set a minimum ionization fraction of 10$^{-8}$ and a minimum neutral gas fraction of 10$^{-4}$ in each cell. 
The simulations include heating and ionization from stars, background UV radiation, and background cosmic rays as described in~\citet{Wall2020}; they also include atomic~\citep{Joung2006}, molecular~\citep{Neufeld1995}, and dust cooling~\citep{Hollenbach1989}. We assume a dust density of 1\% of the gas density when the gas temperature is below $T_{\mathrm{sputter}} =$~3 x 10$^5$~K, and a dust cross-section of 10$^{-21}$ cm$^2$. The gas is assumed to be at solar metallicity, and we do not follow enrichment from winds or SNe. 

\subsection{Stellar dynamics}\label{sec:petar}
We use \textsc{PeTar}~\citep{Wang2020b} to handle collisional stellar dynamics. \textsc{PeTar} combines different approaches to stellar dynamics for different separation regimes. The implementation of \textsc{PeTar} in \textsc{Torch} is presented in~\citet{Polak2024a}, with further discussion in~\citet{Cournoyer-Cloutier2024b} for the coupling with binaries. We set a global maximum time step for our simulations of $dt_{\mathrm{soft}} = 31.25$~yr, set in conjunction with the separations at which each algorithm is used. This allows the simulation time step to vary according to the instantaneous gas conditions while ensuring that the time step remains short enough to resolve the orbits of wide binaries. 

For inter-star distances below the regularization radius $r_{\mathrm{bin}}$, stars are handled by slow-down algorithmic regularization~\citep{Wang2020a}.
For inter-star distances between $r_{\mathrm{bin}}$ and $r_{\mathrm{in}}$, a direct N-body approach is adopted, with a fourth-order Hermite integrator~\citep{Makino1992}. At separations greater than $r_{\mathrm{out}}=10 \, r_{\mathrm{in}}$, gravity between the stars is calculated with a tree code~\citep[\citeauthor{Barnes1986}~\citeyear{Barnes1986}, as implemented by][]{Iwasawa2016}. Between those last two regimes, a weighted average of the direct N-body and tree approaches is adopted~\citep[with weight depending on the distance, see][for the changeover function]{Wang2020b}.
Individual stars can have larger $r_\mathrm{in}$ and $r_\mathrm{out}$ if they are more massive than the average stellar mass in the simulation; their changeover radii are then increased by a factor of $(m_i/\langle m \rangle)^{1/3}$. 
We adopt $r_{\mathrm{bin}} = 50$~au and $r_{\mathrm{out}}= 625$~au (which corresponds to $r_{\mathrm{out}}=12.5 \, r_{bin}$, following the default \textsc{PeTar} ratio). 

\subsection{Binary identification}\label{sec:binaryID}
We identify binaries before every stellar evolution step. We build a tree of neighbors from the three-dimensional positions of the stars, and use it to calculate the pairwise binding energy for each star and its 10 nearest neighbors. The most bound companion, for each star, is saved if its binding energy is positive and the semimajor axis of the corresponding binary is smaller than 10,000 au. 
Stars are allowed to change companions throughout the simulation, and updating the binary pairs every time step allows us to account for modifications to the orbits and exchanges due to few-body encounters. 

\subsection{Stellar and binary evolution}\label{sec:seba}
Our model for binary feedback is designed to be used alongside stellar dynamics, which may modify the binaries' orbits beyond what is expected from the evolution of the system in isolation, or result in exchange interactions involving post-mass transfer systems. Being able to restart from previously evolved stars is therefore a necessary feature.  
%
We use the stellar and binary evolution code \textsc{SeBa}~\citep[][]{PortegiesZwart1996, Nelemans2001, Toonen2012}\footnote{\url{https://github.com/amusecode/SeBa}, commit \texttt{a6f4b64}} to obtain all stellar properties throughout our simulations. 
\textsc{SeBa} is a rapid stellar evolution code that uses the analytical fits of~\citet{Hurley2000} to model the stellar evolution of the stars.
In contrast with other rapid stellar and binary evolution codes, \textsc{SeBa} uses the instantaneous properties of a star -- such as its mass, core mass, and age -- in addition to its zero-age main sequence (ZAMS) properties to calculate its subsequent evolution. 
%
\textsc{SeBa} includes prescriptions for wind mass loss from MS and evolved stars, SN mass loss, SN kicks, binary mass transfer, and binary orbital evolution. 

\subsubsection{Wind mass loss rate}
A star's mass loss rate depends on its current mass, luminosity, metallicity, and surface gravity. We use the default \textsc{SeBa} wind mass loss rates for solar metallicity ($Z_{\odot}$ = 0.02), which are described in detail in~\citet[][their appendix A.1]{Toonen2012}.
The wind mass loss rates for OB MS stars are based on the rates from~\citet{Vink2000, Vink2001} and from~\citet{Nieuwenhuijzen1990}, with a correction factor of 1/3 to account for wind-clumping effects~\citep[][]{Bjorklund2020}. 
%
The mass loss rates for the Hertzsprung gap (HG) stars are identical to the ones for OB MS stars, except in the luminous blue variable (LBV) regime, where a mass loss rate of 1.5 x 10$^{-4}$ M$_{\odot}$ yr$^{-1}$ is 
used~\citep[][]{Belczynski2010}. For helium stars, the mass loss rates are the maximum of rates from~\citet{Reimers1975} and~\citet{Hurley2000}. For giants with a hydrogen envelope, the adopted mass loss rates are the maximum of the (corrected)~\citet{Nieuwenhuijzen1990} rates and of the~\citet{Reimers1975} rates, with the same LBV correction as the HG stars.


\subsubsection{Luminosities}
For each star, we calculate the luminosity in the FUV (5.6--13.6 eV) and ionizing ($>$ 13.6 eV) bands from the stellar radii and surface temperatures calculated in \textsc{SeBa}. We calculate the luminosity assuming blackbody emission for the FUV band. For the ionizing band, we use the atmospheric model of~\citet{Lanz2003} for stars with surface temperatures between 27,500 and 55,000 K, and assume blackbody emission beyond those temperatures.

\subsubsection{Supernovae}
We use the default \textsc{SeBa} prescription for core-collapse SNe, which is itself based on the prescription by~\citet{Fryer2012}. This prescription provides us with the remnant mass and ejecta mass; we describe in Section~\ref{sec:SNinj} the injection of the ejecta in the simulation. For SNe in binaries, we update the velocity of both the star and the remnant self-consistently from the instantaneous change in primary mass, with an additional contribution from the SN's natal kick. As the stellar evolution code has no information about the exact positions of the stars, only their orbital parameters, the resultant kick is calculated for a random time in the binary's orbit. 

\subsubsection{Mass transfer \& common envelope}
The masses and radii of both stars are evolved concurrently in \textsc{SeBa}. \textsc{SeBa} detects MT due to RLOF and the ejection of a common envelope if they occur. The exact criteria to identify those phases of binary evolution are presented in detail in ~\citet[][their appendices A.2 and A.3]{Toonen2012}. 
The two key quantities that we use are the total amount of mass lost by the system during the interaction and the change in semimajor axis due to the interaction. 
For the binary evolution calculations, we use the default \textsc{SeBa} values for $\alpha$ and $\lambda$, which describe the efficiency of the envelope's unbinding from CE evolution, and $\beta$, which describes the MT efficiency.  

\subsubsection{Orbital evolution}
Wind mass loss, conservative and non-conservative MT, CE evolution, and tidal circularization all result in changes to the orbital parameters of a binary. For detached binaries (i.e., binaries in which neither star fills its Roche lobe), mass loss results in the binary becoming less bound, and can even result in the disruption of the binary in the case of rapid mass loss from SNe. 
The fractional increase in semimajor axis due to the mass loss will be approximately equal to the fractional decrease in the binary's total mass~\citep[see][]{Toonen2012}.
Conservative MT results in a decrease in the semimajor axis while $M_1 > M_2$. For non-conservative MT, the semimajor axis may either decrease or increase depending on the mass ratio $M_2/M_1$, the angular momentum imparted to the ejected material, and the MT efficiency~\citep[see][for recent observations and simulations]{Nuijten2025, Rodriguez2025}.
The semimajor axis decreases in the case of a CE phase or under the effects of tidal circularization. 

\subsection{Interaction detection}\label{sec:detection}
We consider a binary to be interacting, for a given time step, if it meets any of the following criteria:
\begin{enumerate}
    \item It is identified by \textsc{SeBa} as a semi-detached or contact binary, or as undergoing CE evolution;
    \item Its binding energy has increased over the stellar evolution step;
    \item Its semimajor axis has decreased over the stellar evolution step;
    \item One of the stars in the binary has accreted mass.
\end{enumerate}
Interactions may be missed by using only criterion (1), as the timescales for MT and CE evolution are much shorter than the time step dictated by cloud-scale hydrodynamics.
We calculate the change in binding energy from 
\begin{equation}\label{eq:dE_bind}
    \Delta E_{\mathrm{orbit}} = \frac{G}{2} \Bigg{(}\frac{M_{1,\mathrm{f}} M_{2,\mathrm{f}}}{a_\mathrm{f}} - \frac{M_{1,0} M_{2,0}}{a_0}\Bigg{)}
\end{equation}
where $M_1$ and $M_2$ denote the primary and companion mass, and $a$ denotes the semimajor axis. The subscript $0$ corresponds to the values before applying stellar and binary evolution, while the subscript f corresponds to the values after stellar and binary evolution. If the stars do not interact, a normal binary evolution step results in wind mass loss from both stars and an increase in semimajor axis, leading to $\Delta E_{\mathrm{orbit}} < 0$.  If the stars interact, the binary may become more bound, leading to $\Delta E_{\mathrm{orbit}} > 0$. This is always the case for CE ejection; we therefore use the CE ejection scheme (Section~\ref{sec:mloss_CE}) if the binary's binding energy has increased. If the binary's binding energy has decreased despite an interaction being detected -- for example in the case of MT resulting in an increase in semimajor axis -- we inject the mass lost from the system as a wind from the star that has lost the most mass (Section~\ref{sec:winds}).

\subsection{Injecting mass loss on the grid}\label{sec:mloss}

Core-collapse SNe, stellar winds, CE ejection, and non-conservative MT all increase the amount of gas, handled with different injection schemes in the code.

\subsubsection{Core-collapse supernovae}\label{sec:SNinj}
The SN prescription corresponds to the default \textsc{Torch} prescription, presented in~\citet{Wall2020}. For each SN, a mixture of kinetic and thermal energy is injected on the grid over the 27 cells closest to the SN, with the fraction of kinetic energy proportional to the background density over the surface area of the nearby cells, following~\citet{Simpson2015}. The SN energy is fixed at 10$^{51}$~ergs, and the amount of mass injected by each SN is calculated from the difference between the stellar mass before the supernova, and the remnant mass. 

\subsubsection{Stellar winds}\label{sec:winds}
We use the wind injection routine described in~\citet{Wall2020}. The wind is injected over a spherical region of radius 3.5$\sqrt{3} \Delta x$, where $\Delta x$ is the cell size at the highest refinement level; all cells over which wind is injected must be at the highest refinement level. The wind material is injected into the ISM with a velocity based on terminal wind velocities derived from observations. We conserve momentum when injecting wind material within the simulation domain. The final gas velocity within a cell in the injection region is
\begin{equation}
    v = \frac{\rho_{\mathrm{w}}v_{\mathrm{w}} + \rho_0 v_0}{\rho_{\mathrm{w}} + \rho_0}
\end{equation}
where $\rho_{\mathrm{w}}$ is the wind density (calculated from the mass loss rate and volume of the injection region), $v_{\mathrm{w}}$ is the wind terminal velocity, and $\rho_0$ and $v_0$ are respectively the background gas density and velocity.

The wind terminal velocities are calculated based on the prescription by~\citet{Kudritzki2000} and~\citet{Vink2000}. The velocity $v_{\mathrm{w}}$ of the injected material depends on the surface temperature $T$ and the effective escape velocity $v_{\mathrm{esc}}$ as
\begin{equation}\label{eq:v_wind}
  v_{\mathrm{w}} =
    \begin{cases}
      v_{\mathrm{esc}} & T < 10^4 \text{ K}\\
      1.4 \, v_{\mathrm{esc}} & 10^4 \text{ K} < T < 2.1 \text{ x } 10^4 \text{ K}\\
      2.65 \, v_{\mathrm{esc}} & T \geq 2.1 \text{ x } 10^4 \text{ K}.
    \end{cases}       
\end{equation}

The effective escape velocity is calculated from
\begin{equation}
    v_{\mathrm{esc}} = \sqrt{\frac{2GM(1-\Gamma_\mathrm{e})}{R}}
\end{equation}
where $M$ and $R$ are the star's current mass and radius. The Eddington ratio $\Gamma_{\mathrm{e}}$ depends on the ratio of radiation pressure to gravity, as
\begin{equation}
    \Gamma_\mathrm{e} = \frac{L\sigma_\mathrm{e}}{4 \pi c G M}
\end{equation}
where $L$ and $M$ are the star's current luminosity and mass, and $\sigma_\mathrm{e}$ is the electron scattering cross-section. We cap $\Gamma_{\mathrm{e}}$ at 0.8. Recent samples of non-interacting O stars~\citep{Bestenlehner2014, Bestenlehner2020, Brands2022} find values of $\Gamma_{\mathrm{e}} \lesssim 0.8$ for MS and Wolf-Rayet stars, and this value is sufficiently high to capture the transition into the optically-thick wind regime~\citep[][and references therein]{Vink2022}. We implement this limit to ensure that stars out of thermodynamical equilibrium due to binary interactions do not get assigned unphysically low wind velocities. 

\subsubsection{Common envelope ejection and non-conservative mass transfer}\label{sec:mloss_CE}

As discussed in Section~\ref{sec:detection}, the simulation time step may frequently be longer than the evolutionary timescales for MT and CE, which prevents us from clearly distinguishing those cases at runtime. We therefore adopt an approach which separates the interacting systems into those for which the binding energy of the binary has decreased over the time step, and those for which the binding energy of the binary has increased. The first group will be composed only of systems having undergone MT from RLOF, while the second may contain both systems that have undergone RLOF or CE ejection. 

Massive binaries are more likely to undergo non-conservative MT than a CE phase~\citep[see, e.g.,][]{Pavlovskii2017}; for those undergoing MT via RLOF, an increase in orbital period is common~\citep[e.g.][]{Lechien2025, Nuijten2025}. Most massive interacting binaries in our simulations therefore show a decrease in orbital energy over their mass transfer phase. We treat mass loss from such systems as a wind from the donor star, and inject it with a velocity calculated with Equation~\ref{eq:v_wind}.

A smaller fraction of systems show an increase in orbital energy. For those systems, we adopt an approach based on the energy formalism~\citep{VandeHeuvel1976, Webbink1984} for CE ejection,
%
%
which is often used to describe the efficiency of the envelope's ejection. Under the energy formalism, binary evolution codes calculate the change in orbital energy $\Delta E_{\mathrm{orbit}}$ following CE ejection from
\begin{equation}\label{eq:alpha}
    E_{\mathrm{bind}} = \alpha \Delta E_{\mathrm{orbit}},
\end{equation}
where $E_{\mathrm{bind}}$ is the envelope's binding energy and $\alpha$ an efficiency parameter. This approach assumes that the energy supplied to unbind material from the star is related to the orbital energy lost by the binary by an efficiency $\alpha$. The presence of other energy sources beyond the gravitational potential energy~\citep[such as H recombination energy, see e.g.][]{Ivanova2018} may result in $\alpha > 1$. The exact value of $\alpha$ is highly uncertain, and may vary as a function of stellar mass.
Compilations of observations and simulations~\citep{Iaconi2019, Roepke2023} yield values from $\lesssim$ 0.1 to $\gtrsim$ 1 for low to intermediate masses. For massive stars, accounting for convection~\citep{Wilson2022} and radiation pressure support in the envelope~\citep{Lau2022} however yields $\alpha\approx1$.

Population synthesis studies adopt a fixed value of $\alpha$, with a common choice being $\alpha=1$~\citep[e.g.][]{Hurley2002, Fragos2023}, and use this value in Equation~\ref{eq:alpha} to calculate the change in orbital separation following CE ejection.  
We invert this approach to calculate the ejecta velocity by using the change in orbital separation to calculate the energy imparted to the ejecta. 
The compiled value of the binary evolution parameters within \textsc{SeBa} remain set to their default values and we assume that the energy supplied to the ejecta is supplied as kinetic energy. We set the ejecta velocity to 
\begin{equation}{\label{eq:vterm_binary}}
    v = \Bigg{(}\frac{2 v_{\mathrm{CE}} \Delta E_{\mathrm{orbit}}}{M_{\mathrm{lost}}}\Bigg{)}^{\nicefrac{1}{2}}
\end{equation}
where $M_{\mathrm{lost}}$ is the binaries' mass loss over the time step and $v_{\mathrm{CE}}$ is treated as a velocity scaling for the ejecta. We adopt a value of $v_{\mathrm{CE}}=1$ (i.e. all the energy supplied to unbind the envelope is injected as kinetic energy for $\alpha=1$), which yields ejecta velocity on the order of the stars' wind velocity. The material is injected over a spherical region of radius 3.5$\sqrt{3} \Delta x$, following the wind injection routine. 

\section{Isolated binaries}\label{sec:MT}

Here, we present simulations of feedback from interacting binaries, which we compare to simulations in which the same two massive stars are evolved using a single star stellar evolution scheme. For those comparisons, the same initial binary properties -- i.e., the masses of both stars, the semimajor axis, and the eccentricity -- are used. The \textsc{Torch} simulations are initialized just before the onset of MT, which then changes the properties of both stars and, subsequently, of the nearby ISM. 
We outline the initial conditions for the gas background medium, then describe the impact of conservative and non-conservative MT and CE ejection on the nearby gas for our tests problems. 

\subsection{Gas initial conditions}

\begin{table*}[t!]
    \centering
    \begin{tabular}{ccccccccccccccccccc}
        \hline
        Name & BE & $v_{\mathrm{CE}}$ & $\Sigma$ & $M_{1, 0}$ & $M_{2, 0}$ & $a_0$ & $e_0$ & $M_{1, \mathrm{f}}$ & $M_{2, \mathrm{f}}$ & $a_\mathrm{f}$ & $e_\mathrm{f}$ & $L_\mathrm{FUV}$  & $L_\mathrm{ion}$ &  $r_{\mathrm{ion}}$ \\
        & & & (M$_{\odot}$ pc$^{-2}$) & (M$_{\odot}$) & (M$_{\odot}$) & (au) & & (M$_{\odot}$) & (M$_{\odot}$) & (au) & & (L$_{\odot}$) & (L$_{\odot}$) &  (pc) \\
        (1) & (2) & (3) & (4) & (5) & (6) & (7) & (8) & (9) & (10) & (11) & (12) & (13) & (14) & (15) \\
        \hline
        \hline
        MT-LD-B & $\checkmark$ & 1 & 10$^{2}$ & 148 & 128 & 2.87 & 0.37 & 30.3 & 108 & 10.5 & 0 & 3.89E5 & 9.56E5 & 0.98 \\
        MT-LD-S & & - & 10$^{2}$ & 148 & 128 & 2.87 & 0.37 & 69.4 & 69.0 & 5.75 & 0.37 & 3.12E5 & 2.16E3 & 0.29 \\
        MT-HD-B & $\checkmark$ & 1 & 10$^{3}$ & 148 & 128 & 2.87 & 0.37 & 30.2 & 108 & 10.6 & 0 & 4.01E5 & 9.54E5 & 0.23 \\
        MT-HD-S & & - & 10$^{3}$ & 148 & 128 & 2.87 & 0.37 & 69.4 & 69.0 & 5.75 & 0.37 & 3.11E5 & 2.16E3 & 0.16 \\
        \hline
        nMT-LD-B & $\checkmark$ & 1 & 10$^{2}$ & 148 & 96 & 18.7 & 0.64 & 30.2 & 71.9 & 14.6 & 0 & 3.97E5 & 8.95E5 & 0.22 \\
        nMT-LD-S & & - & 10$^{2}$ & 148 & 96 & 18.7 & 0.64 & 68.8 & 71.6 & 32.7 & 0.65 & 3.12E5 & 3.93E3 & 0.38 \\
        nMT-HD-B & $\checkmark$ & 1 & 10$^{3}$ & 148 & 96 & 18.7 & 0.64 & 30.2 & 71.9 & 14.4 & 0 & 6.10E5 & 9.27E3 & 0.05 \\
        nMT-HD-S & & - & 10$^{3}$ & 148 & 96 & 18.7 & 0.64 & 68.8 & 71.6 & 32.8 & 0.65 & 3.13E5 & 3.93E3 & 0.21 \\
        \hline
        CE-LD-B & $\checkmark$ & 1 & 10$^{2}$ & 148 & 10 & 5.00 & 0.40 & 29.7 & 10.0 & 0.22 & 0 & 9.64E3 & 4.66E5 & 0.21 \\
        \hline
    \end{tabular}
    \caption{Overview of the isolated binaries' simulations. All simulations are conducted in a uniform cubic box with side $L$~=~4~pc, with an initial temperature of 100 K and no initial turbulence. Columns (5) to (8) correspond to the ZAMS values while columns (9) to (15) correspond to values 1~kyr after the mass transfer event. Columns: (1) simulation label, (2) use of binary stellar evolution, (3) velocity parameter for common envelope ejection, (4) initial gas surface density of the background medium, (5) ZAMS primary mass, (6) ZAMS companion mass, (7) initial semimajor axis, (8) initial eccentricity, (9) final primary mass, (10) final companion mass, (11) final semimajor axis, (12) final eccentricity, (13) final FUV luminosity, (14) final ionizing luminosity, (15) final size of the \ion{H}{2} region.}
    \label{tab:runs}
\end{table*}

The isolated binary simulations are conducted in a box of size $L$~=~4~pc. We use a uniform background density without turbulence and an initial temperature of 100 K. There is no initial magnetic field. We  refine up to a refinement level of 4, which corresponds to a resolution of 3.125 x 10$^{-2}$ pc. We use two different background densities to mimic different star-forming environments. We use surface densities of 10$^2$ and 10$^3$ M$_{\odot}$ pc$^{-2}$, similar to surface densities typical of giant molecular clouds in the disks of star-forming spiral galaxies and in starburst galaxies~\citep[see, e.g.][]{Sun2018}. 
We present an overview of the test simulations in Table~\ref{tab:runs}. The run names starting with MT correspond to conservative MT simulations, those starting with nMT correspond to non-conservative MT simulations, and those starting with CE correspond to CE ejection simulations. LD and HD respectively denote the low and high density background medium. The final letter in the run label corresponds to the stellar evolution scheme used: B denotes binary stellar evolution and S denotes single star stellar evolution.

\subsection{Conservative mass transfer}
The first test is a case of conservative MT. This changes the temperature, luminosity, and mass loss rates of both stars, but does not result in the injection of additional material into the ISM from the mass transfer event. The properties of the binary and the gas 1~kyr after the mass transfer event are presented in Table~\ref{tab:runs}, along with the ZAMS properties of the binary.
We evolve the system in \textsc{SeBa} for 3.285 Myr (i.e. just before the primary moves off the MS) before placing the stars within the simulation box. The evolved stars have masses $M_1$~=~69.42~M$_{\odot}$ and $M_2$~=~69.46~M$_{\odot}$, and the semimajor axis has increased to 5.72 au due to the wind mass loss. For convenience, we will continue to refer to the initially most massive star -- i.e. the donor star -- as the primary, despite the other star being more massive at late times. 

\subsubsection{Stellar and binary evolution}

Between the start of the simulation and the onset of mass transfer, the primary evolves into a HG star, and its radius increases rapidly.  
In the simulations with binary stellar evolution, this increase in radius leads to RLOF: material is stripped from the primary and accreted by the companion.
During the MT event, which lasts $\lesssim$~100~yr, approximately 39~M$_{\odot}$ of material is removed from the primary and accreted by the companion. The orbit circularizes and the semi-major axis increases to 10.5~au. 
The primary becomes a stripped star, exposing its helium core, while the companion is rejuvenated by the accretion of the primary's envelope.
This has an important impact on the FUV (5.6 -- 13.6 eV) and ionizing ($\geq 13.6$ eV) luminosity of the stars. We report those values in Table~\ref{tab:runs} for the stars 1~kyr after the MT event. Accounting for binary evolution increases both the FUV luminosity and the ionizing luminosity, but has the largest effect on the ionizing luminosity, which increases by almost three orders of magnitude for this system. 
The total system mass is not directly affected by the mass transfer event, as the MT is conservative; the $\sim$0.2 M$_{\odot}$ difference in the final total masses is due to the differences in the wind mass loss rates of the post-MT stars compared to the stars evolved with single star stellar evolution. 

\subsubsection{Gas properties}

\begin{figure}[htb!]
    \centering
    \includegraphics[width=0.895\linewidth, clip=True, trim=3cm 1cm 0cm 1cm]{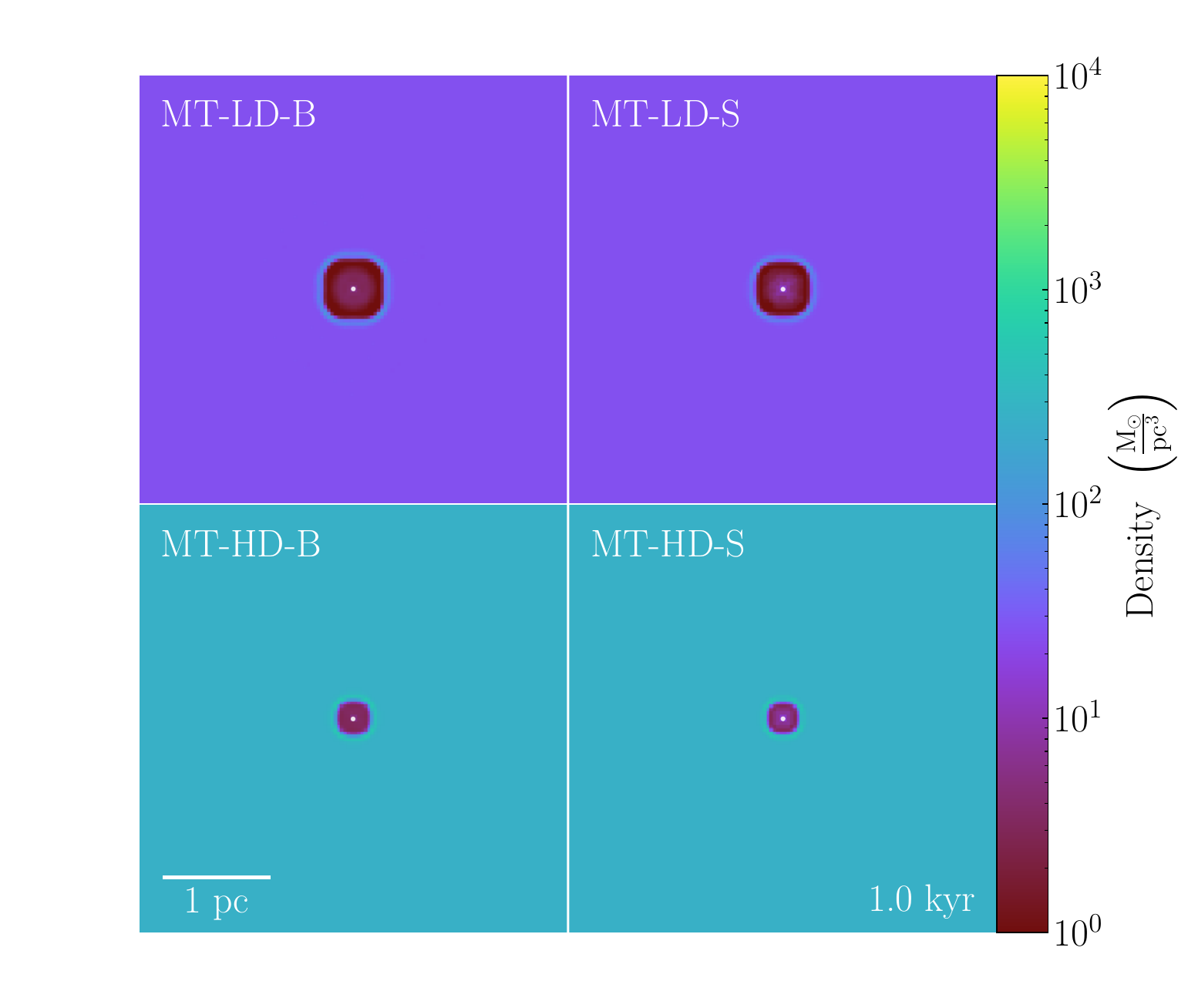}
    \includegraphics[width=0.895\linewidth, clip=True, trim=3cm 1cm 0cm 1cm]{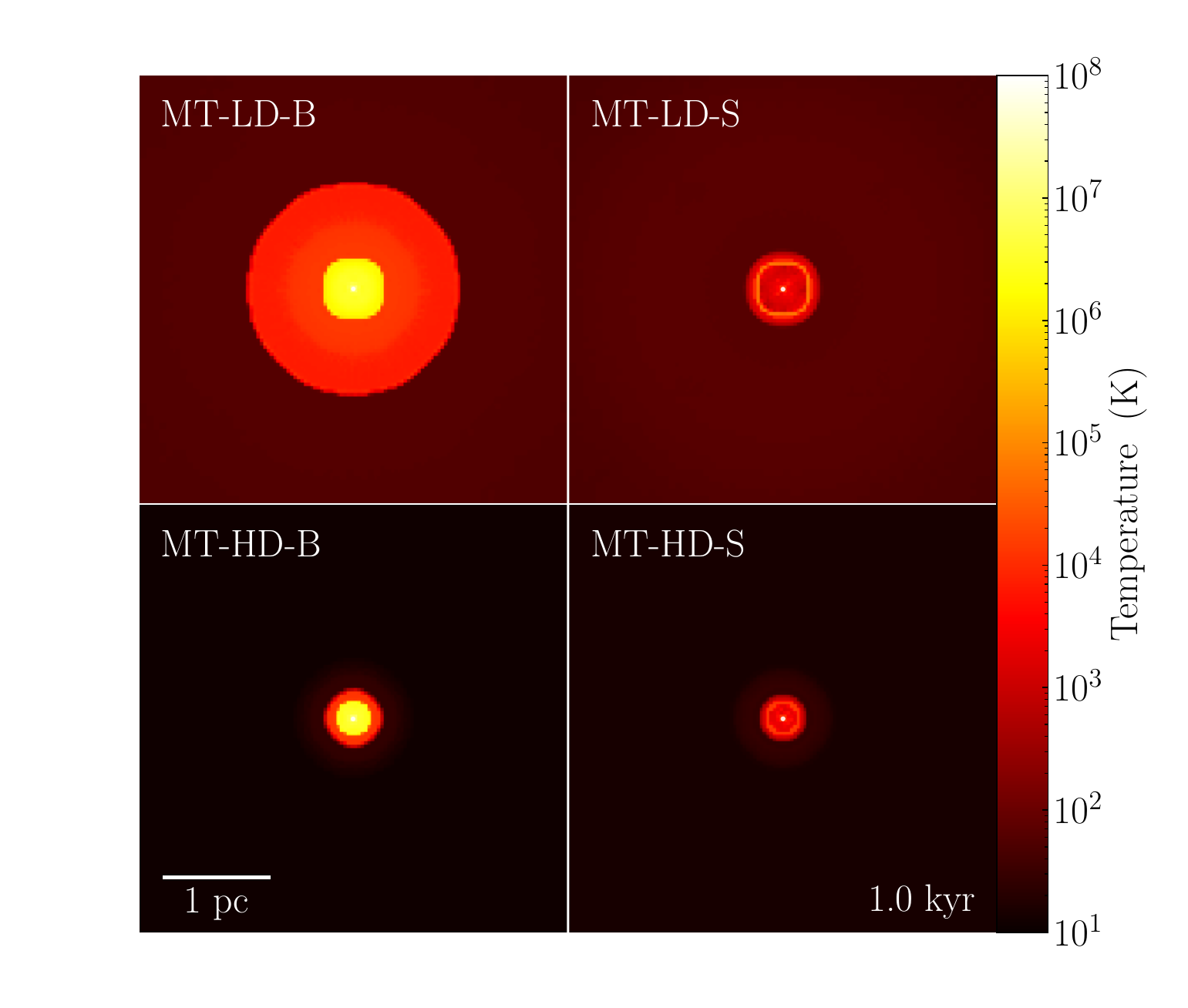}
    \includegraphics[width=0.895\linewidth, clip=True, trim=3cm 1cm 0cm 1cm]{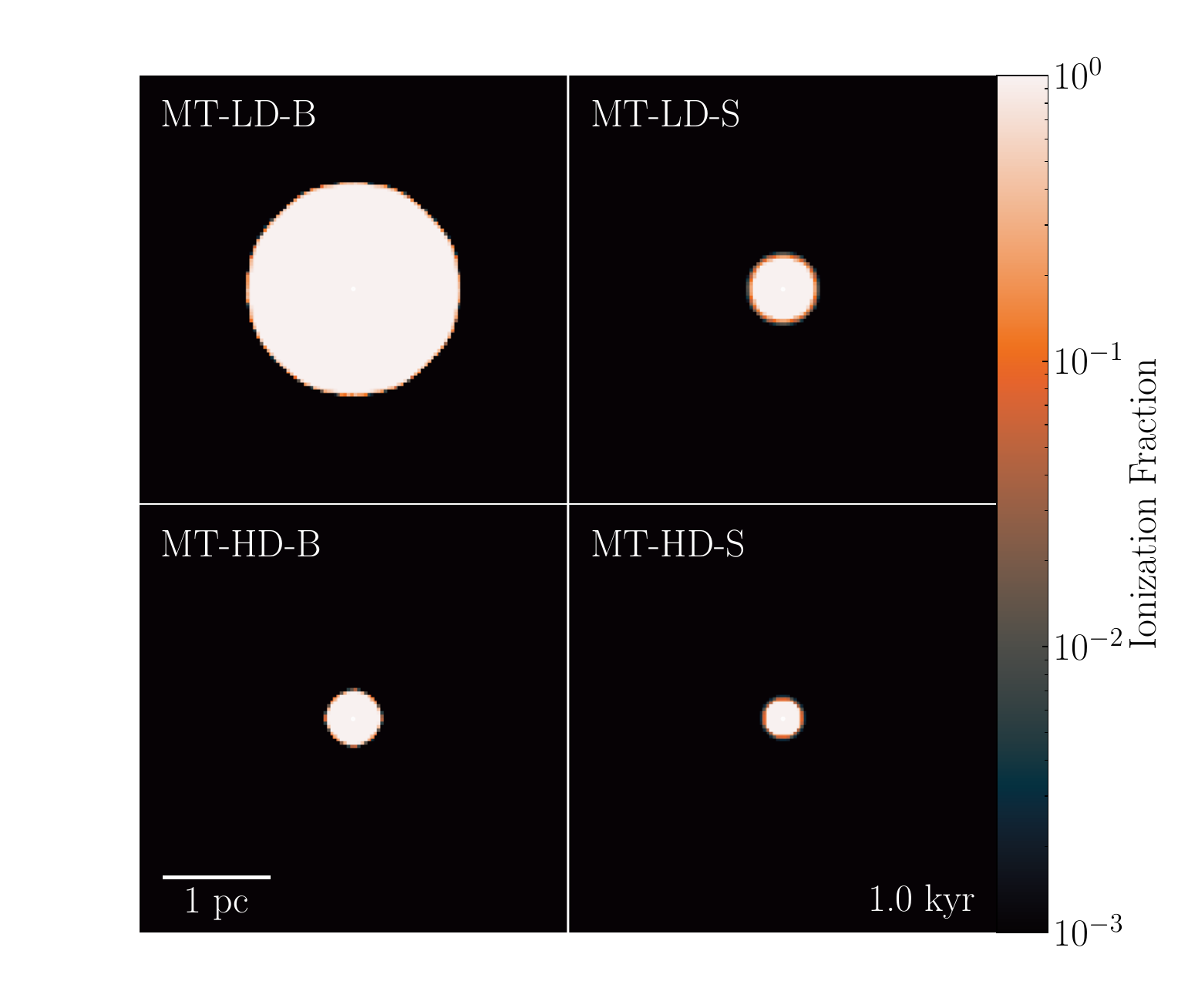}
    \caption{Density (top), temperature (middle) and ionization fraction (bottom) in the midplane for the conservative MT simulations, 1~kyr after the mass transfer event. For each panel, the simulations on the left correspond to runs with binary stellar evolution, and those on the right to runs with single star stellar evolution. }
    \label{fig:MT-slices}
\end{figure}

We present in Figure~\ref{fig:MT-slices} the density, temperature, and ionization fraction in the midplane, 1~kyr after the mass transfer event, for all four MT simulations.
Including the effects of MT results in a larger bubble with a hotter wind, and a larger \ion{H}{2} region. 
The differences are more subtle in the higher density medium, but the \ion{H}{2} region is larger in the simulation with binary evolution, and the temperature within the \ion{H}{2} region is higher.
Those changes are driven by the change in surface temperature of the stars due to the primary's RLOF. 

\begin{figure*}[htb!]
    \centering
    \includegraphics[width=\linewidth, clip=True, trim=2cm 1cm 2cm 2.3cm]{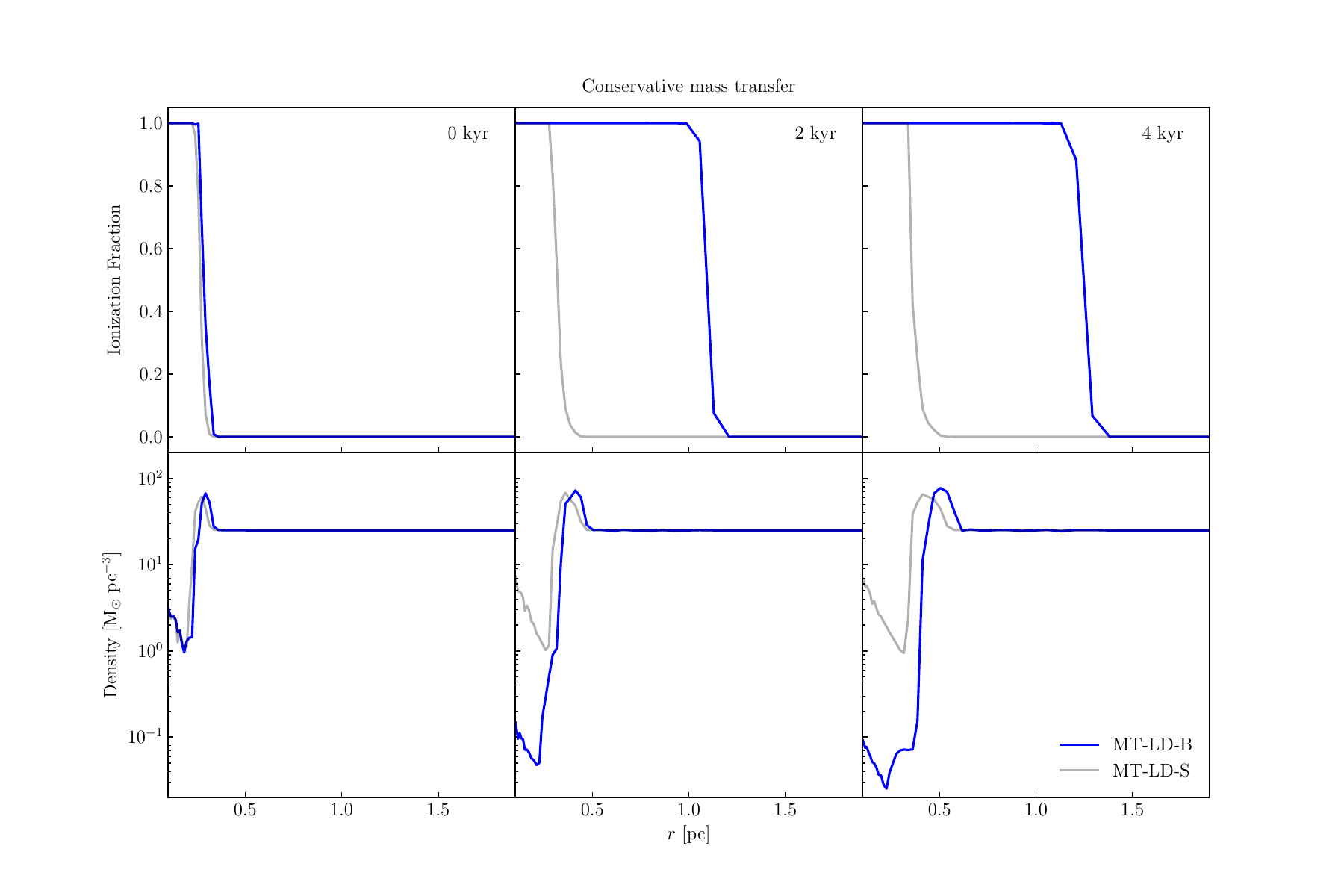}
    \caption{Ionization fraction (top) and density (bottom) as a function of radial distance to the binary's center of mass, for the conservative MT simulations in the lower density medium, 0 kyr, 2kyr and 4 kyr after the mass transfer event. Each radial bin is the mass-weighted average of the gas properties at this radial distance from the binary's center of mass. The blue line corresponds to the simulation with binary stellar evolution and the grey line to the simulation with single star stellar evolution.}
    \label{fig:MT-radial}
\end{figure*}

In Figure~\ref{fig:MT-radial}, we plot the ionization fraction and the density as a function of distance to the system's center of mass, for the simulations with the lower gas surface density, MT-LD-B and MT-LD-S. Both simulations show an \ion{H}{2} region around the binary and develop wind bubbles with a shock front; the wind bubble is also larger in the simulation with binary evolution, although the difference is more subtle. 
We obtain the \ion{H}{2} region radius from the volume of gas with an ionization fraction above 99\%; we then assume that the \ion{H}{2} region is spherical and calculate the equivalent radius as 
\begin{equation}
    r_{\mathrm{ion}} = \Bigg{(} \frac{3 V_{\mathrm{ion}}}{4 \pi} \Bigg{)}^{1/3}
    \label{eq:radius}
\end{equation}
where $V_{\mathrm{ion}}$ is the volume of ionized gas and $r_{\mathrm{ion}}$ the calculated equivalent radius.
We report it in Table~\ref{tab:runs}. 

\subsection{Non-conservative mass transfer}\label{sec:uMT}
The second test case we simulate is non-conservative MT: the primary fills its Roche lobe and starts transferring mass onto its companion, but the mass cannot be fully accreted and some is ejected from the system. This changes the temperature, luminosity, and mass loss rates of both stars, and results in a very high mass loss rate during the MT event. 
We evolve the binary with \textsc{SeBa} for 3.285 Myr before placing it in the simulation; the stars have masses $M_1$~=~69.41~M$_{\odot}$ and $M_2$~=~71.73~M$_{\odot}$, and are still on the MS when we start the simulation, while the semimajor axis has increased to 32.3 au due to wind mass loss.
%
The properties of the binary and the gas 1~kyr after the MT event are also presented in Table~\ref{tab:runs}, along with the binary's ZAMS properties.



\subsubsection{Stellar and binary evolution}

The primary evolves into a HG star just before the onset of mass transfer. During the MT event, which lasts $\sim 200$ yr, $\sim 38$ M$_{\odot}$ of material is removed from the primary. The companion however only successfully accretes $\sim$0.3 M$_{\odot}$, resulting in an instantaneous mass loss rate of $\sim$ 0.2 M$_{\odot}$ yr$^{-1}$ for the MT event. This is about three orders of magnitude higher than the mass loss rates of the evolved stars in the single star stellar evolution scheme.
The helium core of the primary is also exposed, which results in an increase of the binary's ionizing luminosity by up to two and a half orders of magnitude and a more modest increase in the binary's FUV luminosity. 
The semimajor axis decreases to $\sim 14.5$ au following the MT event, while the orbit becomes more circular. 
In contrast with the conservative MT runs, where the evolution of the binary itself was almost identical in the low and high density background media, the non-conservative mass transfer runs show variations in the final semimajor axis $a_{\mathrm{f}}$ and eccentricity $e_{\mathrm{f}}$ between the runs. Those differences arise from the gravitational attraction of the ejecta on the stars and -- by affecting the orbital evolution of the binary and therefore the MT event -- lead to differences in the post-MT radii and temperatures of the stars, and therefore in the final FUV and ionizing luminosities of the binary.

\subsubsection{Gas properties}

\begin{figure*}[t!]
    \centering
    \begin{minipage}{.5\textwidth}
        \centering
        \includegraphics[width=\linewidth, clip=True, trim=6cm 1cm 4cm 1cm]{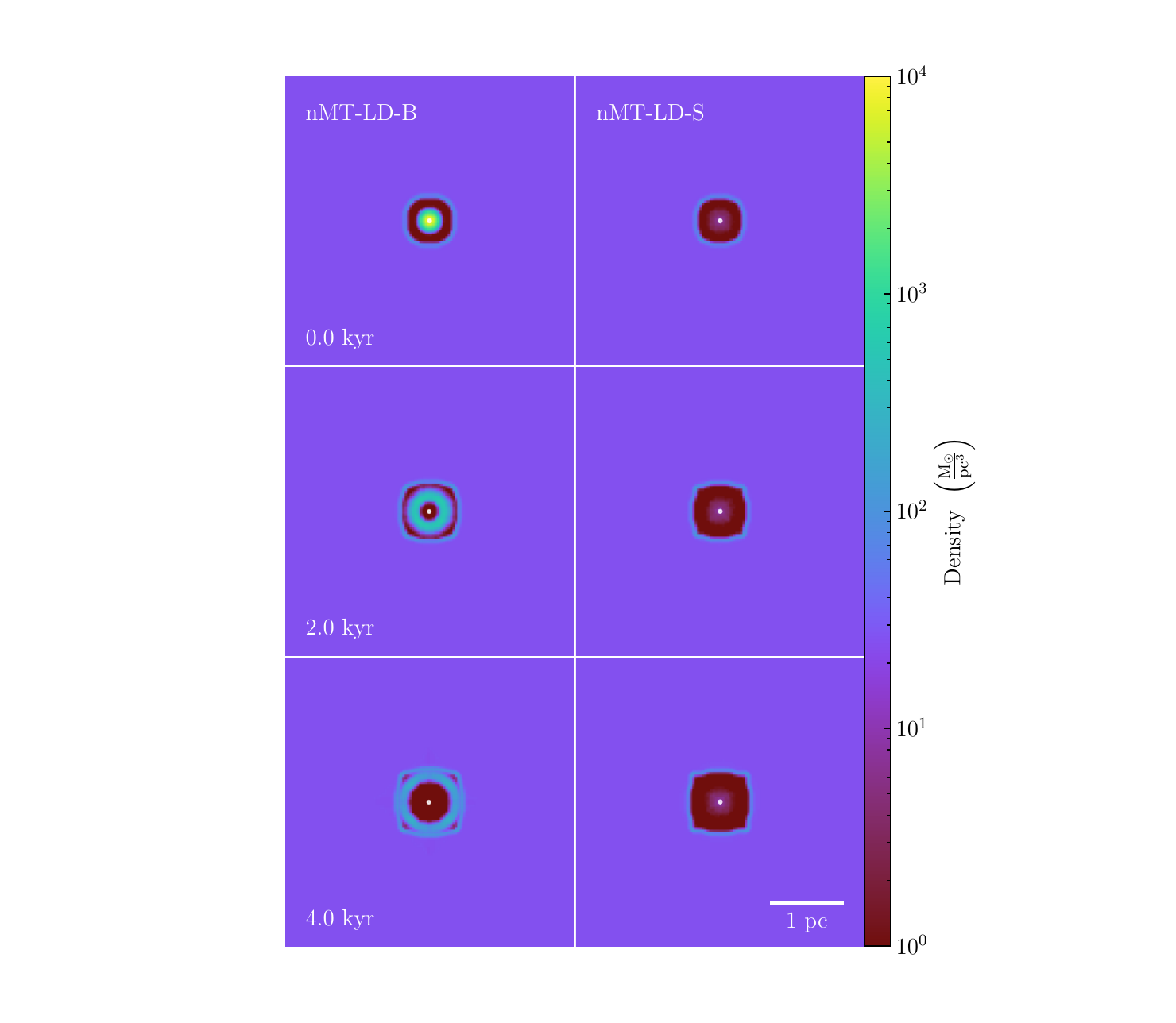}
    \end{minipage}%
    \begin{minipage}{0.5\textwidth}
        \centering
        \includegraphics[width=\linewidth, clip=True, trim=6cm 1cm 4cm 1cm]{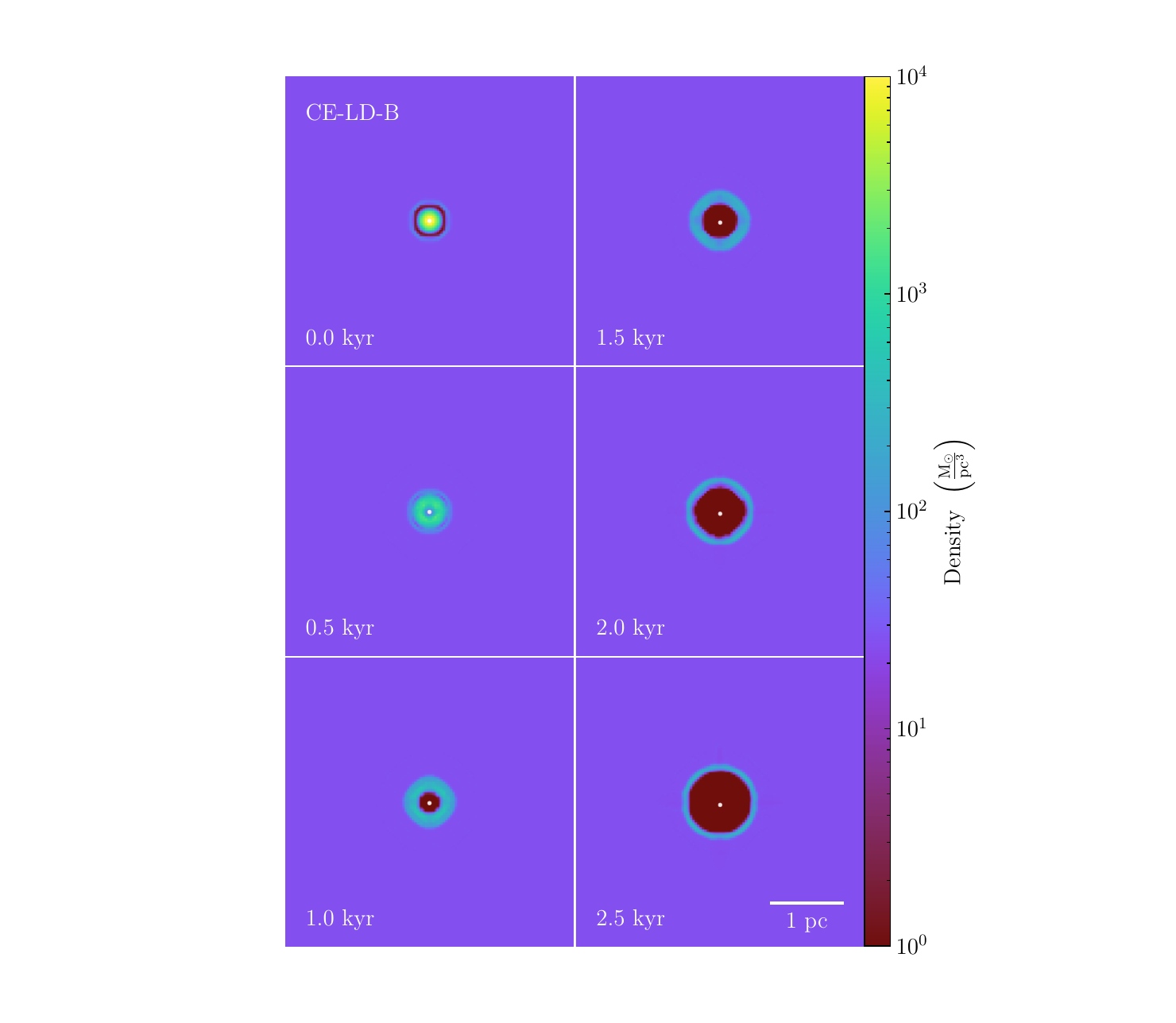}
    \end{minipage}
    \centering
    \caption{The left panel shows density slices in the midplane for the non-conservative MT simulations immediately, 2 kyr, and 4 kyr after the mass transfer event. The right panel shows density slices in the midplane for the CE ejection simulation immediately after the envelope is ejected, then in 0.5 kyr increments until 2.5 kyr. All simulations shown are conducted in the lower density medium.}
    \label{fig:nMT-slices}
\end{figure*}

\begin{figure*}[p!]
    \centering
    \includegraphics[width=0.95\linewidth, clip=True, trim=2cm 1cm 2cm 2.3cm]{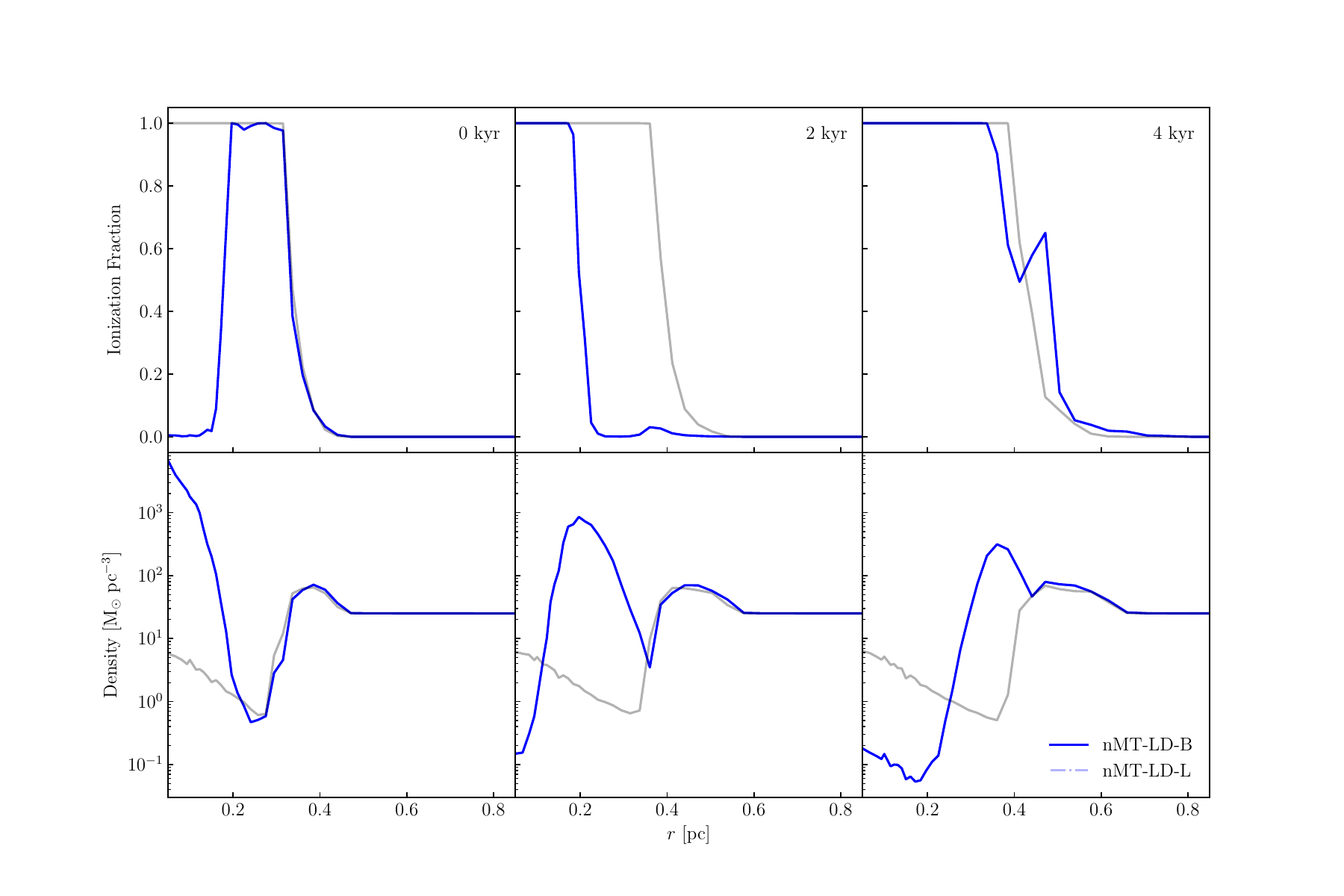}
    \includegraphics[width=0.95\linewidth, clip=True, trim=2cm 1.1cm 2cm 2.3cm]{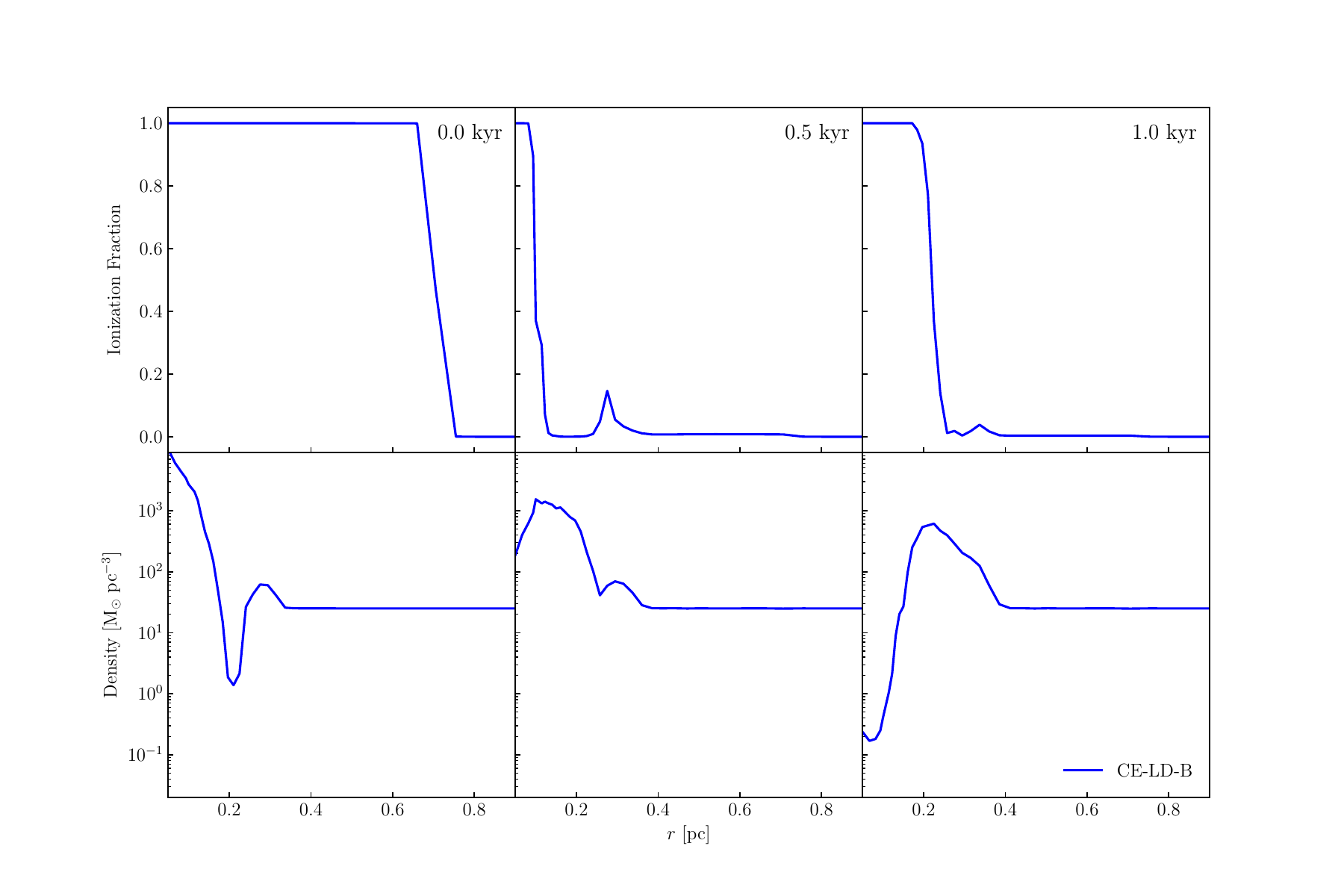}
    \caption{Ionization fraction (top row) and density (bottom row) as a function of radius for the non-conservative MT (top panel) and CE (bottom panel) simulations, at the same times as in Figure~\ref{fig:nMT-slices}. Each radial bin is the mass-weighted average of the gas properties at this radial distance from the binary's center of mass. The solid blue line corresponds to the simulation with binary stellar evolution and the grey line (in the top panel) to the simulation with single star stellar evolution.}
    \label{fig:nMT-radial}
\end{figure*}

We present in the left panel of Figure~\ref{fig:nMT-slices} the density in the midplane immediately after, 2 kyr, and 4 kyr after the non-conservative MT event, for the lower gas background density. We plot the radial ionization fraction and density profiles in Figure~\ref{fig:nMT-radial}. During the early stages of MT, the binary becomes more bound, and the ejecta are injected with a velocity calculated from the change in energy. Most of the mass is injected using the wind routine, however, because despite the orbit shrinking due to mass transfer, the increase in binding energy from the change in semimajor axis is not sufficient to offset the decrease in binding energy from the mass loss, resulting in $\Delta E_{\mathrm{orbit}} < 0$ (Equation~\ref{eq:dE_bind}). The inner, thicker shell -- which contains most of the ejecta -- however moves faster than the outer shell and eventually sweeps it up. Although the ejecta from non-conservative mass transfer is injected using the wind velocity, and therefore the same velocity as the outer shell, the gas velocity in each cell is calculated from momentum conservation. Larger ejecta masses therefore result in larger changes to the gas velocity around the stars.
The loss of stellar material from the non-conservative MT event results in a smaller ionization region right after the MT event. As the ejecta are heated and moves away from the star, the ionization region grows again, reaching a size similar to that of the single star stellar evolution case within 4 kyr. 

The case of non-conservative MT involves not only changes to the stars themselves but also changes to their environment. This is illustrated by the gas density profiles in the top panel of Figure~\ref{fig:nMT-radial}, where developing a clear shock front is not instantaneous. In this case as well, the behavior of the binaries in the gas background differs from what is predicted by standalone \textsc{SeBa} binary evolution simulations. As mass transfer takes place over several time steps, the behavior of the binary after the onset of MT is influenced by the presence of the ejecta and its effects on the local gravitational potential. The region closest to the binary is dominated by the ejecta rather than the background medium; in this case, $\sim$38~M$_{\odot}$ of stellar material is injected within a radius $r \lesssim 0.2$ pc. 
This results in differences in the semimajor axes and luminosities for the simulations with different background gas densities, as reported in Table~\ref{tab:runs}. Those differences are however much smaller than the differences between the runs with binary stellar evolution and without binary stellar evolution. 

\subsection{Common envelope ejection}
We also present test simulations of CE ejection. The binary consists of a 10 M$_{\odot}$ black hole orbiting a star of initially 148 M$_{\odot}$, with a semimajor axis of 5 au and an eccentricity of 0.4. The system has a small mass ratio $M_2/M_1$ at the time of RLOF, leading to CE ejection~\citep{Pavlovskii2017}. 
We use $v_{\mathrm{CE}}=1$ to calculate the ejecta velocity with Equation~\ref{eq:vterm_binary}. Due to the instantaneous nature of CE ejection in our simulations, the background density does not influence the evolution of the system; we therefore only present one background density. The properties of the binary and the gas 1~kyr after the mass transfer event are also reported in Table~\ref{tab:runs}. 

\subsubsection{Stellar and binary evolution}
We evolve the primary for 3.285 Myr before placing it in the simulation. At the start of the simulation, the primary has a mass $M_1$~=~69.41~M$_{\odot}$ and is still on the MS. It evolves into a HG star just before the CE phase. The CE ejection removes almost 40 M$_{\odot}$ of material from the primary, circularizes the orbit, and decreases the semimajor axis to 0.22 au. 

\subsubsection{Gas properties}

We present the density  in the midplane immediately after the envelope is ejected, then in 0.5 kyr increments until 2.5 kyr in the right panel of Figure~\ref{fig:nMT-slices}. 
The ejecta forms a thick shell around the star. The velocity of the ejecta is sufficient to clear out an ionized, low-density bubble around the star within 1~kyr. We plot the radial ionization fraction and density profiles in the bottom panel of Figure~\ref{fig:nMT-radial}.
Although the gas near the star is ionized at the time of CE ejection, the dense ejecta cool quickly, so that the gas close to the star is not fully ionized at early times. CE ejection also increases the gas density close to the star. The CE simulation nonetheless forms an \ion{H}{2} region within 1~kyr.
The calculated ejecta velocities for $v_{\mathrm{CE}}=1$ are at most a few hundreds of km s$^{-1}$, which results in slower ejecta than fast O-star winds with feedback bubbles that retain the overall behavior of a normal \ion{H}{2} region.

\section{Demonstration problem: Cluster of massive binaries}\label{sec:cluster}

We present a suite of simulations of small groups of massive binaries, to showcase the  simultaneous handling of stellar dynamics and stellar evolution. Each simulation contains 10 massive binaries (20 massive stars).
and are run for 50 kyr. 
We present the initial conditions for the gas background medium and the stars (Section~\ref{sec:cluster ics}), the selection of the binaries (Section~\ref{sec:sampling}), and the evolution of the stars and binaries during the simulations (Section~\ref{sec:cluster se}).
The results from the cluster simulations are presented then discussed in Sections~\ref{sec:overview}, \ref{sec:dynamics} and \ref{sec:HII}. 

\begin{center}
\begin{table*}[tb!]
    \centering
    \begin{tabular}{cccccccccccc}
        \hline
        \,\, Name \,\, & \,\, BE \,\, & \,\, Stars \,\, & \,\, $\Sigma$ \,\, & \,\, $\Delta M$ \,\, & \,\, $L_{\mathrm{FUV}}$ \,\, & \,\, $L_{\mathrm{ion}}$ \,\, & \,\, $N_{\mathrm{str}}$ \,\, & \,\, $N_{\mathrm{acc}}$ \,\, & \,\, $N_{\mathrm{a} \downarrow}$ \,\, & \,\, $N_{\mathrm{a} \uparrow}$ \,\, & \,\, $r_{\mathrm{ion}}$ \,\, \\ 
        & & & (M$_{\odot}$ pc$^{-2}$) & (M$_{\odot}$) & (L$_{\odot}$) & (L$_{\odot}$) & & & & & (pc)\\
        (1) & (2) & (3) & (4) & (5) & (6) & (7) & (8) & (9) & (10) & (11) & (12) \\ 
        \hline
        \hline
        C1-LD-B & $\checkmark$ & Set 1 & 10$^{2}$ & 297 & 1.19E6 & 2.88E6 & 7 & 0 & 6 & 2 & 9.69 \\ 
        C1-LD-S & & Set 1 & 10$^{2}$ & 16.4 & 2.49E6 & 1.23E6 & - & - & - & - & 2.93 \\ 
        C1-HD-B & $\checkmark$ & Set 1 & 10$^{3}$ & 89.5 & 2.67E6 & 2.12E6 & 2 & 0 & 2 & 0 & 1.67 \\ 
        C1-HD-S & & Set 1 & 10$^{3}$ & 16.4 & 2.49E6 & 1.23E6 & - & - & - & - & 1.42 \\ 
        \hline
        C2-LD-B & $\checkmark$ & Set 2 & 10$^2$ & 172 & 2.71E6 & 1.01E7 & 12 & 8 & 1 & 9 & 6.40 \\ 
        C2-LD-S & & Set 2 & 10$^{2}$ & 93.4 & 1.04E6 & 7.25E3 & - & - & - & - & 1.32 \\ 
        C2-HD-B & $\checkmark$ & Set 2 & 10$^{3}$ & 137 & 2.71E6 & 1.04E7 & 13 & 7 & 1 & 9 & 2.90 \\ 
        C2-HD-S & & Set 2 & 10$^{3}$ & 93.4 & 1.04E6 & 7.25E3 & - & - & - & - & 0.76 \\ 
        \hline
        C3-LD-B & $\checkmark$ & Set 3 & 10$^{2}$ & 149 & 3.45E6 & 1.19E7 & 10 & 10 & 1 & 9 & 9.23 \\ 
        C3-LD-S & & Set 3 & 10$^{2}$  & 76.4 & 1.72E6 & 1.31E4 & - & - & - & - & 1.94 \\ 
        C3-HD-B & $\checkmark$ & Set 3 & 10$^{3}$ & 56.4 & 3.63E6 & 1.23E7 & 9 & 9 & 0 & 8 & 2.42 \\ 
        C3-HD-S & & Set 3 & 10$^{3}$  & 76.4 & 1.73E6 & 1.31E4 & - & - & - & - & 1.13 \\ 
        \hline
    \end{tabular}
    \caption{Overview of the cluster simulations. The quantities for columns (5) to (12) are calculated at the end of the simulation, after 50 kyr. Columns: (1) simulation name, (2) use of binary stellar evolution, (3) set of evolved stars, (4) initial gas surface density, (5) mass loss during the simulation, (6) cluster FUV luminosity, (7) cluster ionizing luminosity, (8) number of stripped stars, (9) number of stars rejuvenated by accretion, (10) number of binaries with semimajor axes decreased by mass transfer, (11) number of binaries with semimajor axes increased by mass transfer, (12) equivalent radius of the ionized region. All the stars are still in binaries at the end of the runs, in their original pairings.}
    \label{tab:cluster-runs}
\end{table*}
\end{center}

\subsection{Initial conditions}\label{sec:cluster ics}
We select the positions and velocities of the binary centers of mass from the positions of the most massive stars in a recent cluster formation simulations~\citep[][]{Cournoyer-Cloutier2024b}. This cluster  has a stellar mass of 2~x~10$^{5}$ M$_{\odot}$ and a binary fraction $> 90\%$ for massive stars. We use the center of mass information for ten massive systems clustered together near the cluster center. The masses of both stars in each binary are selected following the procedure outlined in Section~\ref{sec:petar}. No other stars beyond those ten binaries are included in the simulations presented in this section. This results in a compact ($\lesssim$~1~pc) configuration of 10 massive binaries.
The high binding energies of the binaries ($>10^{47}$~erg) and their initial slightly infalling velocities ensure that the group of binaries remains bound. Comparable or higher concentrations of massive binaries are observed in the Arches~\citep{Clark2023} and R136~\citep{Sana2013} YMCs. 

The binaries are placed in a uniform medium, within a box of size $L$~=~16~pc. The background is turbulent, with an initial velocity dispersion of 5 km s$^{-1}$. The initial temperature is 100~K and there is no initial magnetic field. We use a refinement level of 4, which corresponds to a finest resolution of 0.125~pc. We use the same initial background gas surface densities of 10$^2$ and 10$^3$ M$_{\odot}$ pc$^{-2}$ as for the isolated binaries. 

\subsection{Binary sampling}\label{sec:sampling}

\begin{figure*}[tb!]
    \centering
    \begin{minipage}{.5\textwidth}
        \centering
        \includegraphics[width=\linewidth, clip=True, trim=0cm 0cm 0cm 0cm]{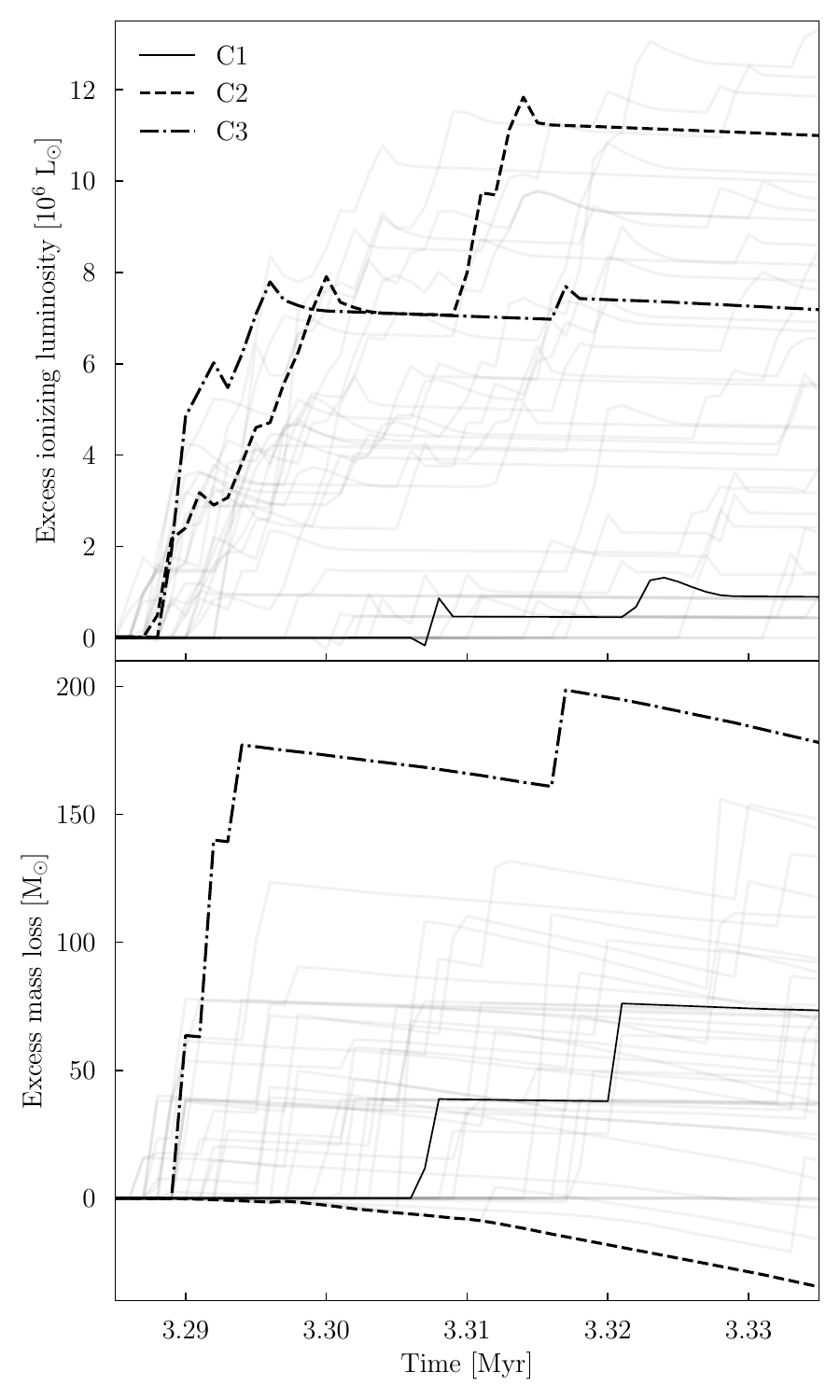}
    \end{minipage}%
    \begin{minipage}{0.5\textwidth}
        \centering
        \includegraphics[width=\linewidth, clip=True, trim=0cm 0cm 0cm 0cm]{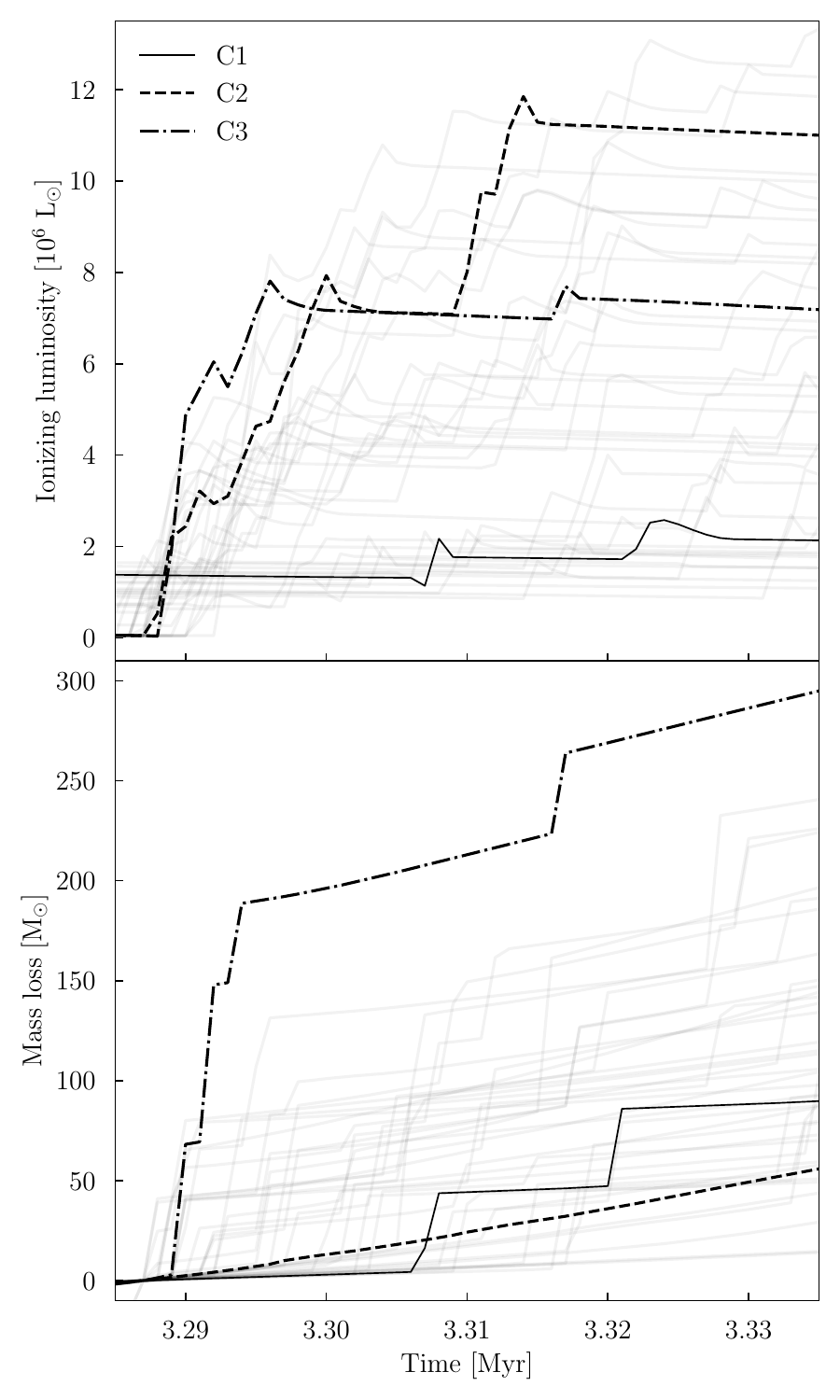}
    \end{minipage}
    \centering
    \caption{Excess (left) and true (right) ionizing luminosity (top) and cumulative mass loss (bottom) from the 40 sampled clusters run with \textsc{SeBa}'s binary stellar evolution scheme. The excess is calculated compared to the same stars evolved with a single star stellar evolution scheme. The black solid, dashed, and dashed-dotted lines (labeled C1, C2, and C3) represent the three clusters we use for the full \textsc{Torch} runs. They are chosen to span the range of ionizing luminosities and mass loss. The fainter lines represent the 37 other clusters for which we only run stellar evolution simulations.}
    \label{fig:ExcessB}
\end{figure*}

Fully sampling a population of binaries requires a stellar mass much higher than a population of single stars
because the companion mass, semimajor axis, and eccentricity of a binary influence its evolution in addition to the primary mass. When accounting for the effects of mass transfer on a stellar population's radiation spectrum, stellar populations with masses up to $\sim 10^6$~M$_{\odot}$ are expected to exhibit significant sample-to-sample variations~\citep{Stanway2023}. Such stellar masses may only be achieved by the most massive YMCs in starburst galaxies~\citep[][]{He2022, Levy2024}, while YMCs in the Local Group have lower masses~\citep[see][]{PortegiesZwart2010}. To estimate the importance of feedback from massive binaries on local YMC formation, it is then crucial to include test cases that do not fully sample the parameter space of close massive binaries.

We run \textsc{Torch} simulations for three different stellar samplings, with two different background gas densities, with and without binary stellar evolution.
We select the clusters for our simulation from the following procedure:
\begin{enumerate}
    \item We sample a~\citet{Kroupa2001} initial mass function (between 0.08~M$_{\odot}$ and 150~M$_{\odot}$) and the binary generation algorithm from~\citet[][based on observations compiled by~\citeauthor{Moe2017}~\citeyear{Moe2017}]{Cournoyer-Cloutier2024b} a total of 40 times (ten times each for clusters of total stellar mass 4~x~10$^4$, 2~x~10$^5$, 1~x~10$^6$, and 5~x~10$^6$~M$_{\odot}$).  The inclinations, eccentric anomalies, longitudes of the ascending node, and arguments of periapsis are sampled from uniform distributions.
    \item For each sampled cluster, we keep the 10 most massive binaries, which dominate the pre-SN feedback budget and provide most of the total radiative feedback due to their very large ionizing luminosities~\citep[][]{Eldridge2022}. 
    We use this set of 10 binaries to represent the clusters' main feedback sources. In the text below, \textit{cluster} refers to this subset of simulated massive stars.
    \item We simulate the 40 clusters in \textsc{SeBa}, with a single star stellar evolution scheme and a binary stellar evolution scheme. 
    \item For each cluster, we compare the mass loss (as a proxy for the strength of the wind feedback) and ionizing luminosity (for radiative feedback) in the \textsc{SeBa} runs with the two different stellar evolution schemes. We present the absolute values of the mass loss and the ionizing luminosity, and the difference in those values between the \textsc{SeBa} runs, in Figure~\ref{fig:ExcessB}. 
    \item We use this comparison to choose three clusters for our full \textsc{Torch} simulations. They are chosen to represent cases with large, medium, and small differences between the single stars and binaries models; the chosen clusters are shown as darker lines in Figure~\ref{fig:ExcessB}.
\end{enumerate}

The simulated clusters 
are labelled C1, C2, and C3. Each cluster is simulated for the two gas background densities, and the two stellar evolution schemes.
C1 is made up of the 10 most massive binaries from a 4~x~10$^4$ M$_{\odot}$ cluster. The binaries cover a fairly large of primary masses (59 -- 138 M$_{\odot}$), semimajor axes (0.95 -- 99 au), mass ratios (0.12 -- 0.67) and eccentricities (0 -- 0.84). C2 and C3 come from 5~x~10$^6$ M$_{\odot}$ clusters, which fully sample the distribution of binary parameters. They have primary masses $\gtrsim 120$ M$_{\odot}$ and mass ratios close to unity, with semimajor axes $\lesssim$~50~au for C2 and $\lesssim$~15~au for C3. The clusters simulated with \textsc{Torch} are selected to encapsulate the full range of excess ionizing luminosity and excess mass loss. From the \textsc{SeBa} simulations, C2 shows the largest difference in the ionizing luminosity compared to single star stellar evolution, and C3 shows the largest difference in the mass loss. 


\subsection{Stellar and binary evolution}\label{sec:cluster se}

We evolve the binaries for 3.285 Myr in \textsc{SeBa} before starting the \textsc{Torch} simulations. All stars are on the MS at that time. We initialize all simulations with the same stellar positions and velocities. The binary center of mass positions and velocities are taken from cluster formation simulations, while the primary and companion relative positions and velocities are taken from the \textsc{SeBa} runs with binary evolution. 
We adopt $v_{\mathrm{CE}} = 1$ for all our cluster runs. 
We report in Table~\ref{tab:cluster-runs} initial and final properties of the simulations. The simulation names use the LD/HD labels for low and high density background, and the B/S labels for binary stellar evolution and single star stellar evolution from the previous section. The first part of the label corresponds to the cluster sampling (C1, C2, or C3). In the following section, we will use the B, S, C1, C2, C3, LD and HD to refer to sets of runs with a shared property.

We report $\Delta M$, the amount of mass lost by the stars during each simulation. For each set of runs, $\Delta M$ is different in the low and high density media for the B runs. $\Delta M$ also differs from the prediction from the \textsc{SeBa} run. This is due to the effects of gravitational dynamics from nearby stars and the background gas. 
In all the B runs except C3-HD-B, $\Delta M$ is larger than in the corresponding S run. This is in contrast with the prediction from pure stellar evolution, in which $\Delta M$ is predicted to be with smaller with binary stellar evolution for the C2 runs. 
%
This mass loss influences the FUV and ionizing luminosities of the clusters. For all clusters, accounting for binary stellar evolution strongly increases the ionizing luminosity: it does so by approximately three orders of magnitude for the C2 and C3 runs, and by a factor of $\sim$2 for the C1 runs. It also increases the FUV luminosity by a factor of $\sim2$ for the C2 and C3 runs, increases it very slightly for C1-HD-B, and decreases it by a factor of $\sim2$ for C1-LD-B.

We report in Table~\ref{tab:cluster-runs} the number of stars stripped (partly or fully) of their envelope via MT or wind self-stripping (following an earlier MT episode), and the number of stars that accreted material via conservative MT. We calculate those values by comparing the masses of each star in the B and S runs. All B simulations show large numbers of stars stripped of their envelopes, with on average almost half of all stars in the simulation losing their envelope. The number of stars that successfully accrete material varies more, ranging from 0 in C1-LD-B and C1-HD-B, to 10 in C3-HD-B. For the clusters that fully sample the distribution of binary orbital parameters (and that we can therefore treat as the proxies for the most massive YMCs), C2 and C3, almost every star has been affected by binary evolution. The orbits of the binaries are also affected by the MT process; for C2 and C3, most of the MT events are conservative, which results in a widening of the orbits once the donor becomes less massive than the accretor. This is not the case for C1 (in particular C1-LD-B), where the MT events are not conservative, and generally lead to a decrease in orbital separation. 

\subsection{Overview of \textsc{Torch} simulations}\label{sec:overview}

\begin{figure*}[phtb!]
    \centering
    \includegraphics[width=\linewidth, clip=True, trim=4.3cm 1cm 2.8cm 1cm]{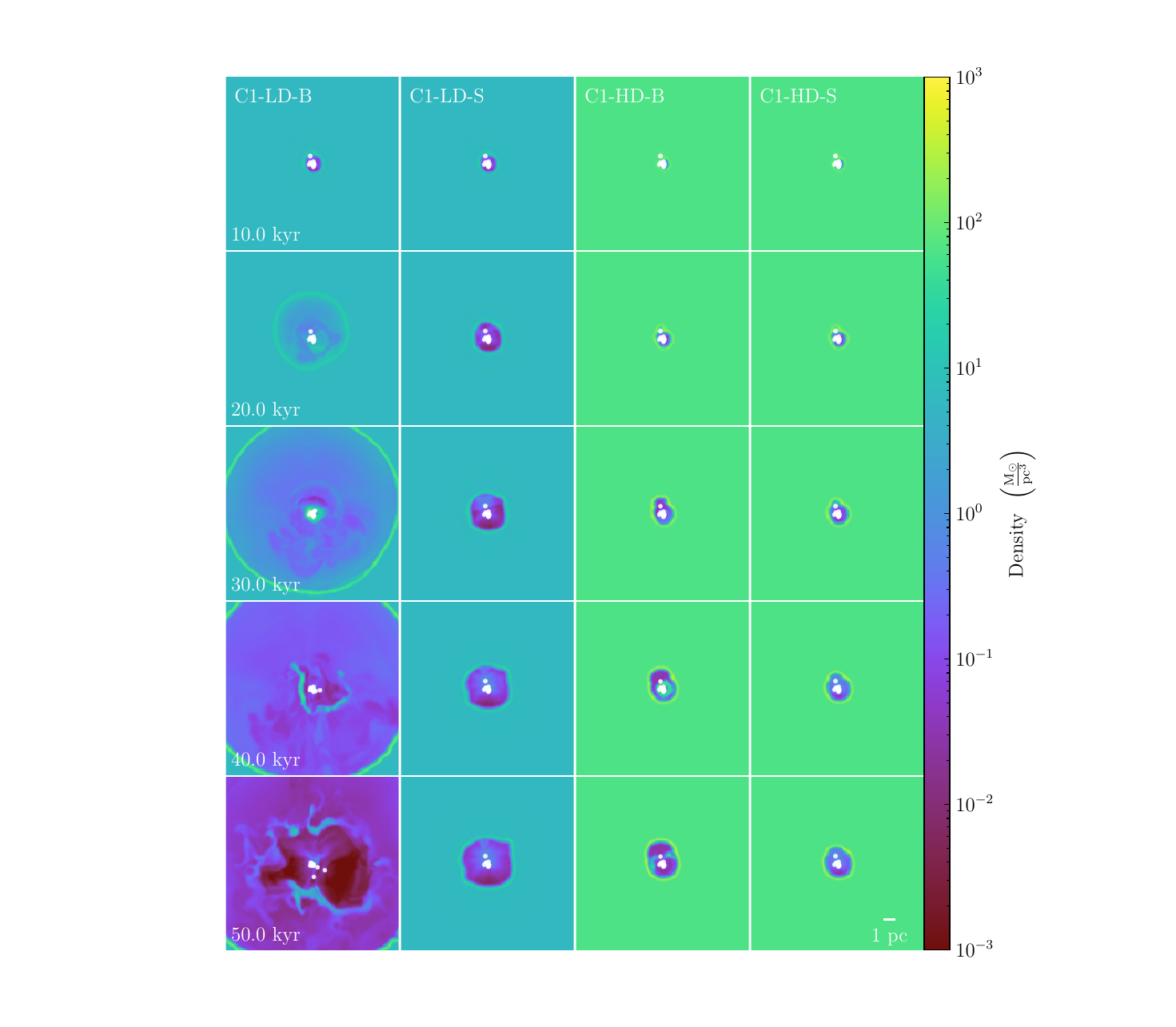}
    \caption{Time evolution of the density in the midplane for the C1 simulations, as a function of time. The B simulations shown in the first and third columns use binary stellar evolution while the S simulations shown in the second and fourth column use single star stellar evolution.}
    \label{fig:C1}
\end{figure*}

\begin{figure*}[phtb!]
    \centering
    \includegraphics[width=\linewidth, clip=True, trim=4.3cm 1cm 2.8cm 1cm]{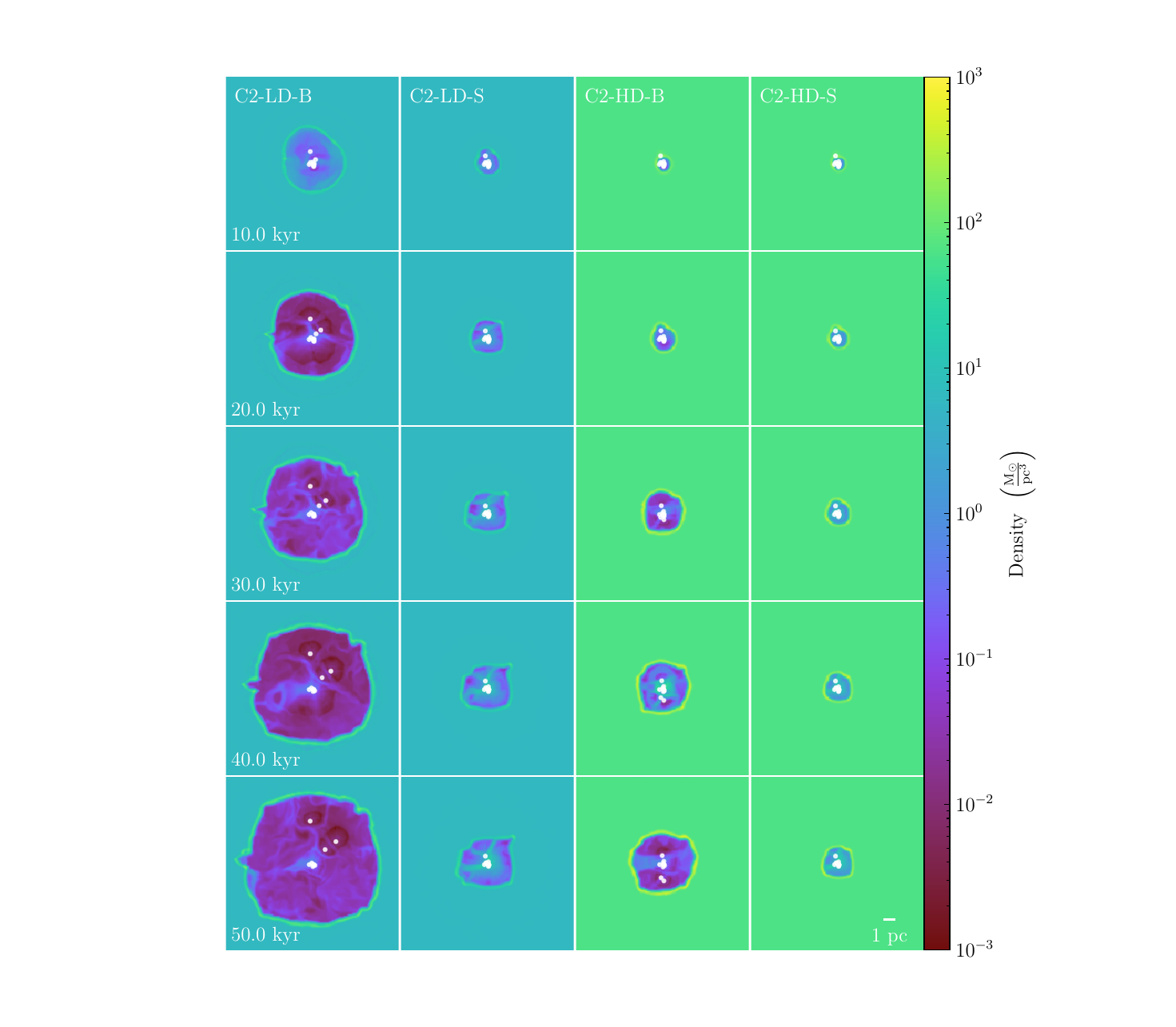}
    \caption{Same as Figure~\ref{fig:C1} for the C2 simulations.}
    \label{fig:C2}
\end{figure*}

\begin{figure*}[phtb!]
    \centering
    \includegraphics[width=\linewidth, clip=True, trim=4.3cm 1cm 2.8cm 1cm]{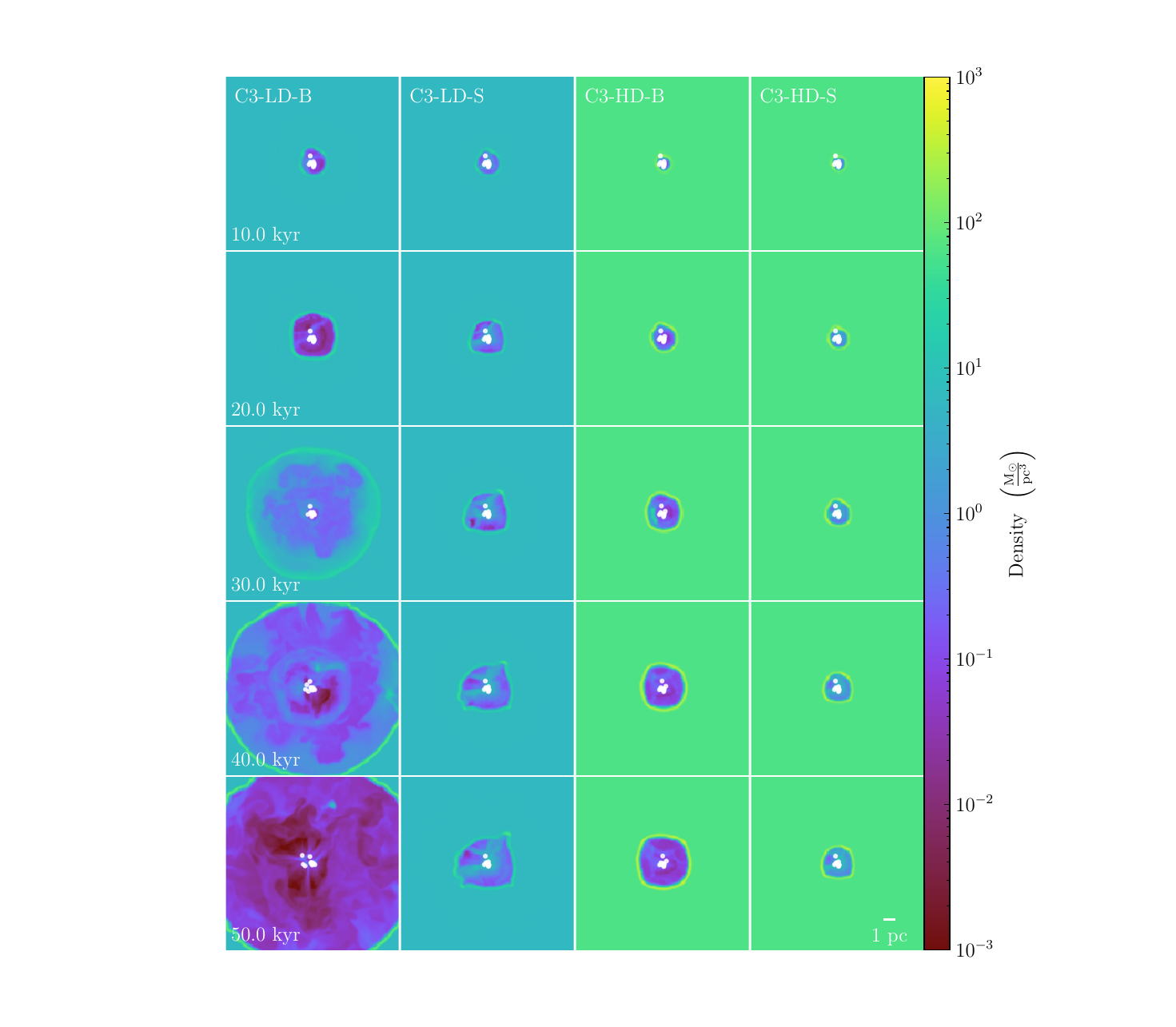}
    \caption{Same as Figure~\ref{fig:C1} for the C3 simulations.}
    \label{fig:C3}
\end{figure*}

We present the density in the midplane every 10 kyr, for the C1, and C2, and C3 runs, in Figures~\ref{fig:C1}, \ref{fig:C2}, and \ref{fig:C3}. Each row shows the four simulations initialized with the same stars, with the two background densities and the two different stellar feedback models. For each pair of simulations with the same stars and the same background gas density, the B simulation forms a larger feedback bubble than the corresponding S simulation. 
The density structure inside the feedback bubbles exhibits a high degree of asymmetry, due to the three-dimensional distribution of the stars. The density substructures in the feedback region are apparent in all runs. 

In the C1 simulations, the differences between the B and S runs are obvious in the four later plots for the LD runs, and the three later plots for the HD runs. The feedback bubble in C1-LD-B is larger than in C1-LD-S, and exhibits more substructured dense gas from the binaries' episodic mass loss. The positions of the stars are also different
which highlights the fact that the different stellar evolution schemes impact the stellar dynamics; we explore this in more detail in Section~\ref{sec:dynamics}. The difference between the sizes of the feedback bubbles is greater in the lower density background medium, with the largest bubble forming in C1-LD-B. 
The difference in the \ion{H}{2} region sizes is more modest for C1-HD-B and C1-HD-S, due to the low number of stars stripped of their envelope by interactions (see Table~\ref{tab:cluster-runs}). 
The three-dimensional configuration of the stars contributes to shaping this feedback bubble. The massive binary above the cluster clears out its surroundings more quickly when binary stellar evolution is used, as illustrated by the 30 kyr snapshot for C1-HD-B and C1-HD-S. This is followed by significant mass loss from the stars below, resulting in the higher gas density around the clustered stars in the 40 kyr snapshot. 

In the C2 simulations, the differences between the B and S simulations are already apparent in the first snapshot shown, after 10 kyr. The main cause of those differences is the rapid start of conservative MT, increasing the ionizing luminosity of the cluster and decreasing the gas density near the stars. In later snapshots, density substructures arise from the different modes of radiative and mechanical feedback coming from individual stars. The stars' positions and velocities differ between C2-LD-B and C2-HD-B, with three fast-moving binaries above the cluster in C2-LD-B and a pair of fast moving binaries in the bottom right quadrant in C2-HD-B. 
Mass transfer lowers the masses not just of the donor stars, but also of some accretors, post-MT. This results in 12/20 stars having lower masses (by $\gtrsim$ 20\%) in C2-LD-B than in C2-LD-S, and 13/20 having lower masses in C2-HD-B than C2-HD-S. 

In the C3 simulations, the size of the feedback bubble also increases with the use of binary stellar evolution, and decreases with increasing background gas density.
In C3-LD-B, all ten primaries successfully transfer mass to their companions, increasing both the FUV and ionizing luminosity. C3-LD-B has lost approximately 75 M$_{\odot}$ in excess of C3-LD-S, in agreement with the prediction from the \textsc{SeBa} simulation without stellar dynamics. C3-HD-B and C3-HD-S are the only pair of simulations in which accounting for binary interactions results in a larger stellar mass by the end of the simulation, in contrast with the expectation from the \textsc{SeBa} simulation, 
which illustrates clearly that the presence of a background gravitational potential -- from the gas and other stars -- may change the outcome predicted by population synthesis studies.

A visual inspection of the gas density, temperature, ionization fraction and pressure in the simulations confirms the large impact of binary stellar evolution on the feedback from star clusters, as predicted by population synthesis studies. The stellar positions and velocities confirm that accounting for binary stellar evolution impacts the dynamics of massive binaries in clustered environments; binary evolution effects should be accounted for when studying the production of massive runaway stars. 
The gravity from the background gas also affects stellar dynamics, and therefore impacts 
individual binaries and their subsequent evolution. 

\subsection{Effects of stellar dynamics}\label{sec:dynamics}

The C1-LD-S and C1-HD-S runs develop small ($\lesssim1\%$) differences in their semimajor axes, which arise from differences in the gas potential due to the different initial gas background densities and subsequent different effects of feedback on that gas. The similar differences in C1-LD-B and C1-HD-B are sufficient to affect the exact timing and stability of mass transfer: the first binary to undergo non-conservative MT does so almost 3 kyr earlier in C1-LD-B than C1-HD-B. Those runs diverge from this point on, leading to the smaller number of stripped stars in C1-HD-B and therefore the more modest increase in feedback bubble size compared to C1-HD-S. 
All stars are still in binaries after 50 kyr in all C1 runs, but
the properties of those binaries have been affected by mass transfer. In C1-LD-B, six systems have shown significant decrease in the semimajor axis following MT, while two systems have seen their semimajor axes increase. In C1-HD-B, only two binaries see their semimajor axes decrease. 

Mass transfer starts at the same time in C2-LD-B and C2-HD-B, and two systems undergo MT within 3 kyr in each simulation. Subsequent MT occurs in C2-LD-B before C2-HD-B, however. In both simulations, mass transfer widens the orbit of the binary in nine cases out of ten. 

In C3-LD-B and C3-HD-B, MT begins within 2 kyr for the two most massive systems. The first difference between C3-LD-B and C3-HD-B arises following a mass transfer episode $\sim25$ kyr after the start of the simulation, when MT widens the orbit in C3-HD-B but tightens it in C3-LD-B. The two runs diverge beyond that point.
The post-MT stellar masses are systematically higher in C3-HD-B than C3-LD-B and most of the semimajor axes (8/9 for mass transfer) are smaller in C3-HD-B than C3-LD-B. 

The cluster simulations all display behavior that diverges from the expectation for population synthesis studies due to the effects of the nearby gas and stars. This can be seen by comparing the $\Delta M$ and luminosity values reported in Table~\ref{tab:cluster-runs} to the values from standalone stellar evolution presented in Figure~\ref{fig:ExcessB}, and by comparing the $\Delta M$ and luminosity values for the pairs of LD and HD runs.
Although the binaries remain bound and remain in their original pairings in all simulations, the contributions of other stars and the background gas to the gravitational potential cause the LD and HD runs to diverge not only in their gas properties -- as expected from the different background densities -- but also in their stellar properties. This is most striking for C1, where the $\Delta M$ is larger by a factor of $\sim3.3$, the ionizing luminosity is larger by a factor of $\sim1.4$, and the FUV luminosity is smaller by a factor of $\sim2.2$ in C1-LD-B compared to C1-HD-B. This is caused by the different number of stripped stars and accretors, which is in turn driven by differences in the semimajor axes of the widest binaries in the simulations. The differences in luminosities are smaller but still present for C2 and C3, which show important differences in $\Delta M$. 
This illustrates the need to model stellar dynamics, hydrodynamics, stellar evolution and stellar feedback in concert.

Another interesting feature is the presence of fast-moving binaries in C1-LD-B (just below and to the right of the cluster), C2-HD-B (below the cluster) and most obviously in C2-LD-B (above the cluster). None of the S runs show this behavior, and runs with the same initial stars but different background densities do not result in the same (or similar) stellar positions and velocities. Although it is not possible to fully disentangle the effects of the background gas from the effects of stellar dynamics (since the first MT event results immediately in changes to both the orbital properties and the properties of the nearby gas), this behavior illustrates clearly that the gravitational potential in which the binaries reside -- shaped by both nearby stars and the background gas -- affects the binaries' evolution. 

\subsection{Expansion of \ion{H}{2} regions}\label{sec:HII}

In addition to the qualitative description of the gas behavior presented in Section~\ref{sec:overview}, we investigate the physical properties of the gas and use them to probe the physical mechanisms driving the expansion of the feedback bubbles in the simulations.

\subsubsection{Size of \ion{H}{2} regions}\label{sec:HII_size}

\begin{figure*}[tb!]
    \centering
    \includegraphics[width=\linewidth, clip=True, trim=0cm 0cm 0cm 0cm]{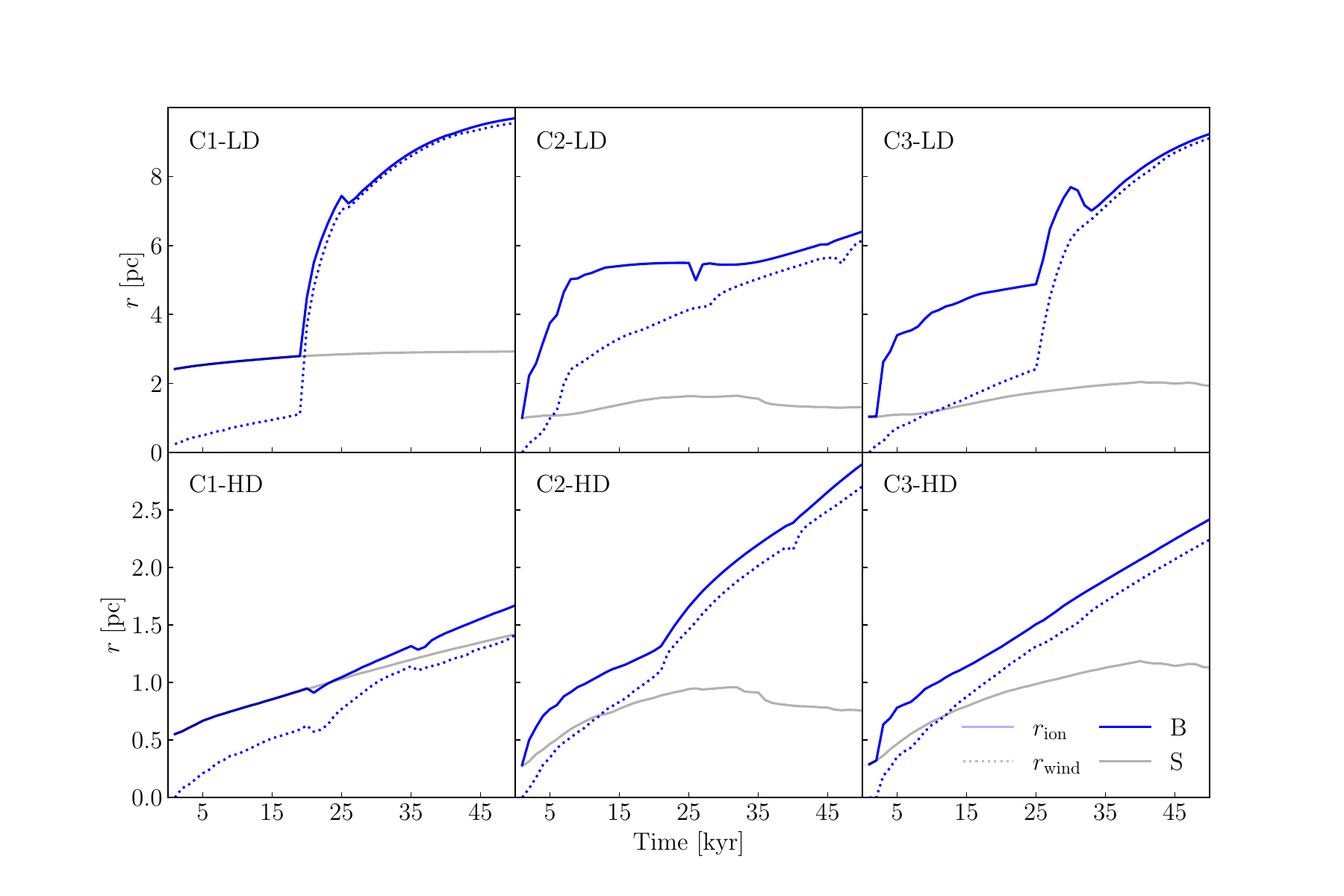}
    \caption{Equivalent radius of the ionized region and shocked wind region (for the B runs) as a function of time, for the cluster runs in the lower (top) and higher (bottom) density medium. The runs with binary stellar evolution are plotted in blue while the runs with single star stellar evolution are shown in grey; the solid lines denote the ionized region and the dotted lines denote the shocked wind region. Dense ejecta from non-conservative mass transfer episodes cause the non-monotonic behavior.}
    \label{fig:Clusters-radii}
\end{figure*}

We calculate the equivalent radius of the \ion{H}{2} region from Equation~\ref{eq:radius}.
For the wind bubble, we follow a similar procedure with the volume of gas with temperature above 10$^6$ K, which corresponds to the shocked wind~\citep[][]{Lancaster2021b}. This temperature cut also traces well the gas with velocity above 200 km s$^{-1}$, separating the ejecta from the background medium. We only calculate the wind radius bubble for the simulations with binary stellar evolution, as the simulations with single star stellar evolution do not form a clear wind bubble with shocked wind, due to the low velocities of the bloated evolved stars' winds. 
%
The sizes of the \ion{H}{2} regions after 50 kyr are summarized in Table~\ref{tab:cluster-runs}, and plotted as a function of time in Figure~\ref{fig:Clusters-radii} (along with the sizes of the wind bubbles for the B runs). The radii for C1-LD-B and C3-LD-B, in which the low-density feedback bubbles reach the edges of the simulation domain, are lower limits. 

All simulations show very clear differences between the B and S runs, with larger \ion{H}{2} regions in the B runs. The smallest difference is for C1-HD-B, in which only two systems undergo mass transfer. For C1, the \ion{H}{2} region radii are equal for the pairs of B and S runs until the onset of non-conservative MT in the first system, after approximately 20 kyr. The radius of the \ion{H}{2} region then increases smoothly due to the increased ionizing radiation from the stripped primary. The non-monotonic behavior of the \ion{H}{2} region and wind radius for C1-HD-B, and of the \ion{H}{2} region for C1-LD-B, are caused by the presence of ejecta from non-conservative MT, which are colder than the bubble interior and therefore decrease the volume of hot ionized gas.

The C2 runs diverge very quickly, following the first MT episode. 
In the high density medium, C2-HD-B exhibits the largest difference with the single star stellar evolution model, in which the gas successfully recombines as the stars cool and move off the MS, leading to a decrease in the ionized volume at late times. Two episodes of non-conservative MT (around 20 and 40 kyr) lead to the two small bumps in the radii for C1-HD-B. 

C3-HD-B follows the smoothest evolution: after the initial MT episode, very early in the simulation, both measures of the size of the feedback bubble keep increasing, reaching values about twice that of the \ion{H}{2} region radius for C3-HD-S after 50 kyr. C3-LD-B shows the most obvious non-monotonic behavior in the expansion of its \ion{H}{2} region. Non-conservative MT just before 30 kyr results in an increase in the gas density near the stars which is apparent from the 2nd and 3rd rows of Figure~\ref{fig:C3}.

\subsubsection{Energetics of \ion{H}{2} regions}

\begin{figure*}[tb!]
    \centering
    \includegraphics[width=\linewidth, clip=True, trim=2.5cm 1.5cm 2.5cm 2.5cm]{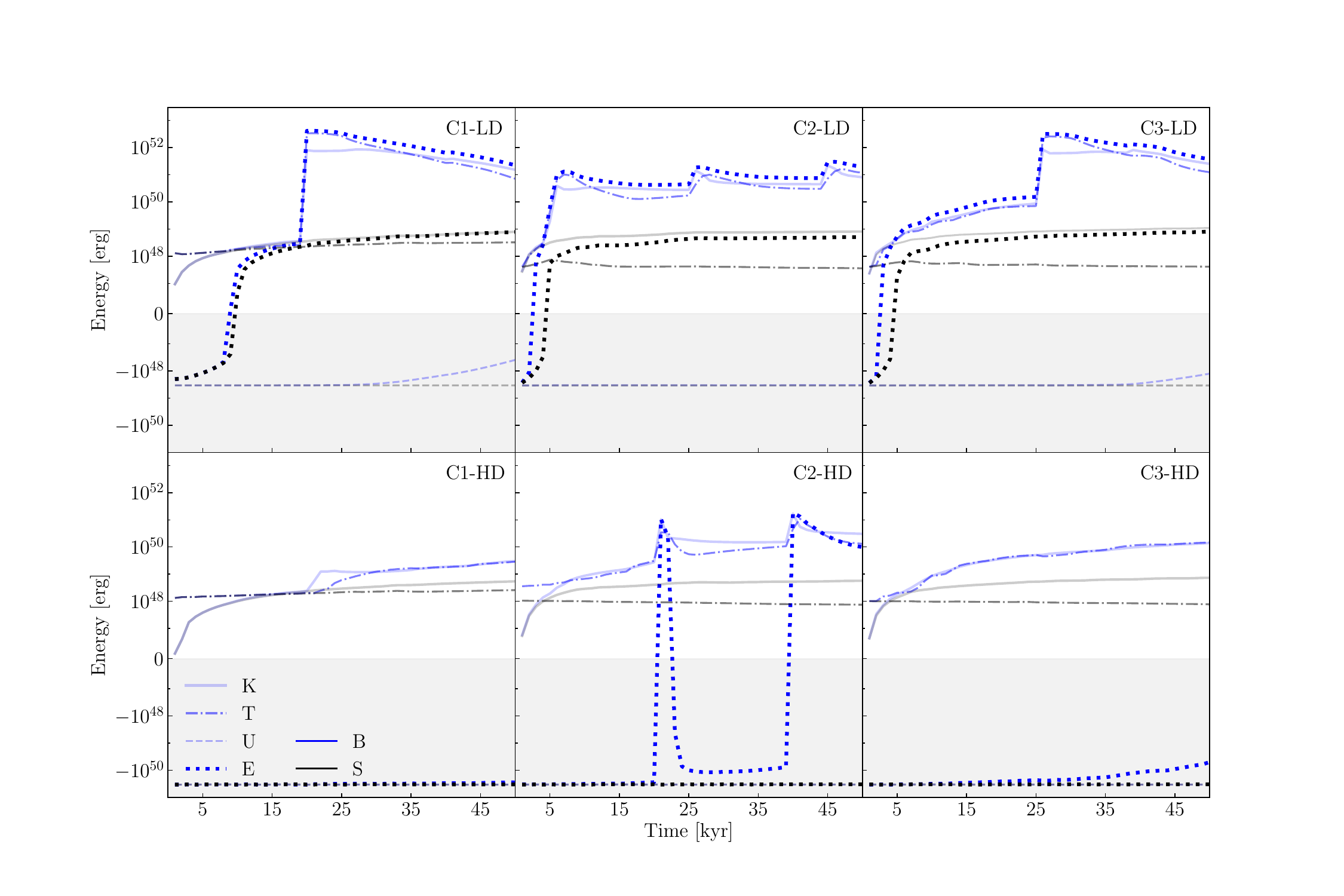}
    \caption{Energetics of the gas in the cluster simulations, as a function of time. Each sub-plot corresponds to a pair of simulations with the same stars and background density (labeled in the top right), with the blue lines corresponding to the runs with binary stellar evolution and the black lines corresponding to the runs with single star stellar evolution. The solid line represents the kinetic energy, the dashed-dotted line represents the thermal energy, the dashed line represents the potential energy and the dotted line represents the total energy. The symlog scale is linear between -10$^{47}$ and 10$^{47}$ ergs. The grey shaded region highlights where the energy is $<$ 0 ergs, i.e. where the gas is globally bound. In all pairs of simulations, the energy is larger with binary stellar evolution.}
    \label{fig:energy}
\end{figure*}

We investigate the energetics of the gas in the simulations, by comparing the kinetic, thermal and gravitational potential of the gas. We calculate the total kinetic energy $K$ of the gas as
\begin{equation}
    K = \frac{1}{2}\sum_{i} m_i |v_i|^2 
    \label{eq:KE}
\end{equation}
where $m_i$ and $v_i$ are the gas mass and velocity in each \textsc{Flash} cell. We calculate the total thermal energy $T$ as 
\begin{equation}
    T = \sum_i \frac{k_\mathrm{B} m_i T_i}{\mu_i m_{\mathrm{H}}(\gamma - 1)}   = \frac{3k_{\mathrm{B}}}{2 m_{\mathrm{H}}} \sum_i \frac{m_i T_i}{\mu_i}
    \label{eq:TE}
\end{equation}
where $k_{\mathrm{B}}$ is the Boltzmann constant, $m_{\mathrm{H}}$ is the mass of an hydrogen atom. We use an adiabatic index $\gamma=5/3$ and calculate the mean molecular weight $\mu_i$ for each cell from the ionization fraction. We use $\mu=0.6$ for fully ionized gas and $\mu=1.3$ for neutral gas, corresponding to gas with solar metallicity. $T_i$ and $m_i$ are the temperature and gas mass in each cell. We calculate the gravitational potential energy of the gas $U$ from
\begin{equation}
    U = \frac{1}{2}\sum_i \Phi_i m_i
    \label{eq:PE}
\end{equation}
where $\Phi_i$ and $m_i$ are the gravitational potential and gas mass in each cell. We also calculate the total energy
\begin{equation}
    E = K + T - U
    \label{eq:E_total}
\end{equation}
and each component of the energy as a function of time, and plot them in Figure~\ref{fig:energy}. In all pairs of simulations, the total energy of the gas is higher when accounting for the effects of binary evolution; the smallest difference is found between C1-HD-B and C1-HD-S, where the kinetic and thermal energies are about one order of magnitude larger in C1-LD-B than in C1-LD-S, but the total energy remains dominated by the potential energy. The evolution of C3-HD-B and C3-HD-S is similar: by the end of the simulation, the total energy is about one order of magnitude less negative in C3-HD-B than C3-HD-S. The largest difference for the HD runs is for C2: accounting for binary stellar evolution successfully unbinds the gas in C2-HD-B while it remains bound in C2-HD-S.
The differences in total energy are large ($\gtrsim$ 2 orders of magnitude) in all the LD runs.

\subsubsection{Sources of pressure in \ion{H}{2} regions}
\begin{figure*}[tb!]
    \centering
    \includegraphics[width=\linewidth, clip=True, trim=2.5cm 1cm 2.5cm 1.5cm]{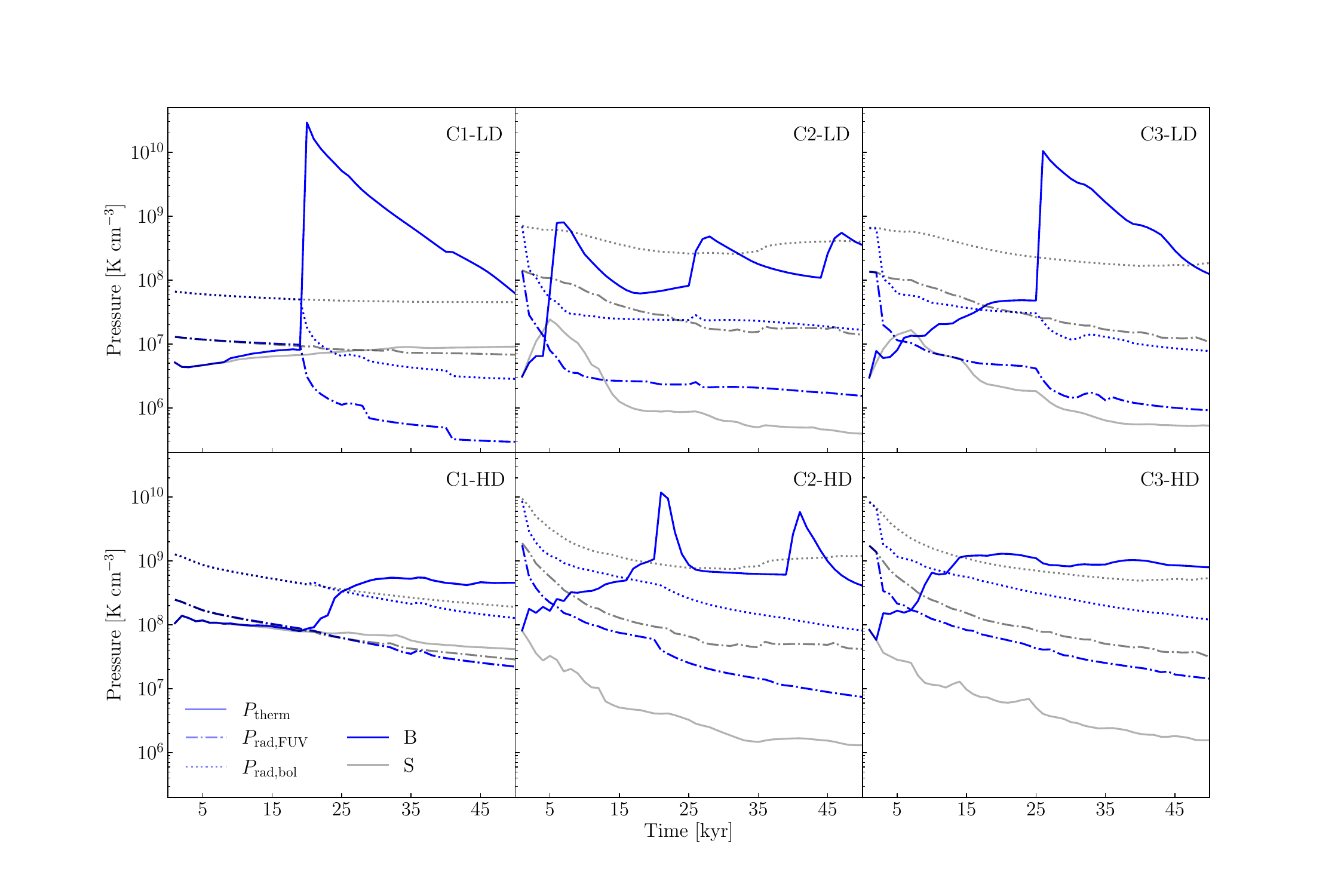}
    \caption{Pressure from radiative feedback as a function of time for the cluster runs. Each column corresponds to a set of stars and each row to a background density. The solid lines denote the thermal pressure from ionized gas (from Equation~\ref{eq:Ptherm}), the dashed-dotted lines denote the direct radiation pressure from FUV radiation (from Equation~\ref{eq:Prad}), and the dotted lines denote the direct radiation pressure calculated from the bolometric luminosity. The very rapid increases in thermal pressure match the timing of mass transfer events, which change the ionizing radiation budget of the cluster on short timescales.}
    \label{fig:pressure}
\end{figure*}

We investigate the time evolution of the pressure within the \ion{H}{2} region surrounding the cluster, to understand what feedback mechanism dominates the evolution of the \ion{H}{2} region.
We evaluate different components of the pressure, following the approach taken in observations of \ion{H}{2} regions~\citep[e.g.][]{Lopez2014, Barnes2020, Barnes2021}. We obtain the thermal pressure $P_{\mathrm{therm}}$ from 
\begin{equation}\label{eq:Ptherm}
    P_{\mathrm{therm}} = \frac{2 \rho T}{\mu m_{\mathrm{H}}}
\end{equation}
where $\rho$ is the gas density, $T$ the gas temperature, and $\mu$ the mean molecular weight. We take the volume average for the ionized gas volume identified in Section~\ref{sec:HII_size}; as the gas is fully ionized, we adopt $\mu=0.6$. 
We also calculate the volume-averaged radiation pressure $P_{\mathrm{rad}}$ from
\begin{equation}\label{eq:Prad}
    P_{\mathrm{rad}} = \frac{3L}{4 \pi r^2_{\mathrm{ion}} c k_{\mathrm{B}}}
\end{equation}
where $L$ is the luminosity of the stars. We calculate the radiation pressure both using the bolometric luminosity (as done in observations) and using only the FUV band (5.6-13.6 eV), as the gas in our simulations is only affected by FUV radiation pressure. We plot those three measures of the pressure in Figure~\ref{fig:pressure}. 
%

%
%

For each pair of simulations, the pressure is higher in the simulation with binary stellar evolution due to the enhanced thermal pressure. This is well in line with the energy analysis presented above, with large increases in gas thermal and kinetic energy due to binary evolution, and significant increases in the volume of hot ionized gas.
A striking example is C1-LD-B, in which the thermal pressure reaches a value almost four orders of magnitude higher than that in C1-LD-S following the onset of mass transfer. Another key difference between the B and S simulations is the presence of several spikes in the time-evolution plot of the thermal pressure. Those maxima are associated with the very high instantaneous mass loss rate from mass transfer events. The first such peak in C1-LD-B and C3-LD-B, after $\sim20$~kyr, matches the rapid increase in \ion{H}{2} region radius (Figure~\ref{fig:Clusters-radii}) and in kinetic and thermal energy (Figure~\ref{fig:energy}). 

Radiation pressure (evaluated from the FUV or bolometric luminosity) is lower in the B runs than the S runs: although the amount of FUV radiation is higher in the B runs than the corresponding S runs (except C1-LD), the inverse square dependence on the radius of the \ion{H}{2} region (which increases due to the increased thermal pressure) dominates, resulting in a decrease in the pressure. In contrast with the B runs, the expansion of the \ion{H}{2} regions in the S runs is driven by radiation pressure rather than thermal pressure. The only potential exception is C1-HD-S, in which the FUV radiation pressure is slightly lower than the thermal pressure (but the bolometric radiation pressure is about one order of magnitude higher).

%

\section{Summary \& Discussion}\label{sec:discussion}

In this paper, we have presented the first implementation of mechanical and radiative feedback from massive interacting binaries for star formation simulations. We have implemented this new feedback module in the cluster formation code \textsc{Torch}, which includes a treatment of magnetohydrodynamics, collisional stellar dynamics, star and binary formation, and stellar feedback. Our new feedback model accounts for the effects of conservative and non-conservative mass transfer (including common envelope ejection) on the mass loss rates, FUV luminosity, and ionizing luminosity of binary interaction products; those changes self-consistently result in changes to the timing of core-collapse SNe. Our feedback module injects the mass lost via non-conservative mass transfer and common envelope ejection in the simulation as a source of mechanical feedback. Our new model also accounts for the gravitational effects of other stars and gas on the binaries, allowing their orbits to be modified both by binary stellar evolution and gravitational dynamics. 

We have tested our new feedback implementation on isolated binaries undergoing conservative and non-conservative mass transfer, as well as common envelope ejection, in Section~\ref{sec:MT}. We have futher presented a suite of simulations of clusters of massive binaries in a turbulent medium, allowing us to demonstrate the effects of the coupling between stellar dynamics and binary stellar evolution. Our key results are as follows:
\begin{enumerate}
    \item As predicted by isolated binary stellar evolution models, accounting for mass transfer increases the ionizing luminosity and the mass loss rate of populations of massive binaries.
    \item Feedback from interacting binaries efficiently couples with the surrounding gas and strongly impacts the nearby ISM. 
    \item Accounting for the effects of binary stellar evolution on stellar feedback increases the size of the \ion{H}{2} regions, increases the kinetic and thermal energy of the surrounding gas, and increases the pressure within the \ion{H}{2} regions.
    \item The expansion of the \ion{H}{2} regions is driven by thermal pressure (rather than radiation pressure) in the presence of interacting binaries.
    \item Stellar dynamics and the gravitational potential of the background gas affect binary stellar evolution by causing changes to the orbits of the binaries. Those changes can affect the timing and efficiency of mass transfer, and therefore the cluster's feedback budget and the evolution of its associated \ion{H}{2} region.
    \item Binary stellar evolution also affects stellar dynamics: accounting for the effects of binary stellar evolution may promote few-body interactions and the ejection of massive stars from their birth environments.
\end{enumerate}

A consistent result across our different simulations -- regardless of the background gas density and the initial stellar sampling -- is that accounting for binary stellar evolution increases the strength of the radiative and mechanical feedback. We conclude that feedback from massive interacting binaries is important to the feedback budget of YMCs of all masses, from the $\gtrsim10^4$~M$_{\odot}$ YMCs observed in the Milky Way~\citep[][]{PortegiesZwart2010} to the $\gtrsim10^6$ M$_{\odot}$ YMCs observed in starburst galaxies~\citep[e.g.][]{He2022, Levy2024}. This additional source of pre-SN feedback may prove crucial in understanding the timescales over which YMCs emerge from their birth environment and stop forming stars.
Although the simulations presented here are a demonstration that feedback from massive interacting binaries has a strong impact on the ISM, they represent an idealized situation.
The effects from more realistic cluster environments may further enhance the impact of feedback from massive interacting binaries.

The simulations were set up to allow Case B MT. MS O-star binaries with orbital periods $\lesssim$ 1000 days~\citep[which represent $\sim$50\% of O stars,][]{Moe2017} can undergo Case A MT if their orbit is sufficiently eccentric. As repeated few-body encounters tend to increase eccentricity~\citep{Heggie1996}, many O stars will undergo Case A MT in dense cluster environments. Using a combination of observations and simulations, \citet{Schneider2014} predict that the first mass transfer event should take place within $\sim2.5$~Myr for 10$^4$~M$_{\odot}$ YMCs, and within $\sim1$~Myr for 10$^5$~M$_{\odot}$ YMCs. The efficiency of feedback from massive interacting binaries in disrupting the gas may also be enhanced by the previous effects of feedback from non-interacting O stars during the earliest stages of cluster formation, which lowers the gas density in the central regions. 

Another potentially important contribution of massive interacting binaries to the feedback budget of galaxies is through the enhanced production of runaway stars. The increased cross-section of the post-interaction systems increase the likelihood that they would undergo few-body interactions ejecting them from the cluster, as has been the case for the fast-moving ($> 30$ km s$^{-1}$) binaries in C1-LD-B, C2-LD-B, and C2-HD-B. 
We expect such ejected binaries to be ubiquitous in massive, dense clusters, in which few-body encounters are frequent. Observations of the rotational velocities of runaway stars around R136~\citep{Sana2022} suggest that a significant fraction of runaways around YMCs are binary interaction products. It is crucial to study concurrently runaway star production and binary interactions in cluster formation models.

Using the new framework presented in this paper in full simulations of star cluster formation is the next logical step. In the meantime, however, simulations that cannot directly include those effects -- due to a lack of (primordial) close binaries or to a stellar evolution scheme that does not include binary evolution -- should still account for the increased feedback budget due to massive interacting binaries. As the expansion of the \ion{H}{2} regions in the runs with binary stellar evolution is driven by the thermal pressure arising from the enhanced ionizing radiation, a simple approach 
is to mimic the effects of non-conservative mass transfer by instantaneously injecting the mass associated with post-MS mass loss as soon as a star leaves the MS, for half of the O stars. Those stripped stars can be evolved as hot Wolf-Rayet stars for the duration of their post-MS lifetimes, increasing the amount of ionizing radiation emitted by the stellar population.  

This first study of massive interacting binaries as a source of radiative and mechanical feedback in cluster-forming regions shows that stellar dynamics, hydrodynamics, and stellar evolution are inter-connected.
Our results clearly demonstrate that binary stellar evolution has a strong effect on a cluster's feedback budget and may play an important role in setting the timescale for gas removal in cluster-forming regions. 
In future studies of star cluster formation, feedback from massive interacting binaries should be considered along with feedback from single stars, hydrodynamics, and collisional stellar dynamics.




\begin{acknowledgments}
C.C.-C. is grateful for the hospitality of the Max Planck Institute for Astrophysics, where this work was started during an extended visit in 2023. C.C.-C. warmly thanks Thorsten Naab for his hospitality during this visit and numerous discussions about stellar feedback and numerical methods. C.C-C. is also grateful to Stephen Justham for several discussions about binary stellar evolution. 

C.C.-C. is supported by a Canada Graduate Scholarship (CGS) -- Doctoral from Natural Science and Engineering Research Council of Canada (NSERC). C.C.-C. also acknowledges funding from a CGS -- Michael Smith Foreign Studies Supplement from NSERC which supported the visit to MPA. A.S. and W.E.H. are supported by NSERC. E.A. acknowledges support from the NASA Astrophysics Theory Program grant 80NSSC24K0935. S.M.A. is supported by a National Science Foundation Astronomy \& Astrophysics Postdoctoral Fellowship under Award No. 2401740. M.-M.M.L. and B.P. acknowledge funding from NSF grant AST23-07950. S.T. acknowledges support from the Netherlands Research Council NWO (VIDI 203.061 grant). 

The code development for this project was done in part on the HPC system Freya at the Max Planck Computing and Data Facility. This research was enabled in part by support provided by Scinet (\url{https://scinethpc.ca/}), Compute Ontario (\url{https://www.computeontario.ca/}) and the Digital Research Alliance of Canada (\url{alliancecan.ca}) via the research allocation FT \#2665: The Formation of Star Clusters in a Galactic Context. This work used Stampede3 at TACC through allocation PHY240335 from the Advanced Cyberinfrastructure Coordination Ecosystem: Services \& Support (ACCESS) program, which is supported by U.S. National Science Foundation grants \#2138259, \#2138286, \#2138307, \#2137603, and \#2138296. 
\end{acknowledgments}

\vspace{5mm}


\software{\textsc{Amuse}~\citep{PortegiesZwart2009, Pelupessy2013, PortegiesZwart2013, PortegiesZwart2019, amuse}, \textsc{Flash}~\citep{Fryxell2000, Dubey2014}, \texttt{matplotlib}~\citep{matplotlib}, \texttt{numpy}~\citep{numpy}, \texttt{scikit-learn}~\citep{scikit-learn}, \textsc{Torch}~\citep{Wall2019, Wall2020}, and \texttt{yt}~\citep{yt}.}





\bibliography{bibliography}{}
\bibliographystyle{aasjournal}



\end{document}